\documentclass[cmp]{svjour}
\usepackage{amsfonts} \usepackage{amssymb} \usepackage{bbm}
\usepackage{graphicx}  
\usepackage{psfrag}    
\usepackage{color}

\def\idty{{\leavevmode\rm 1\mkern -5.4mu I}} 
\def\id{{\rm id}}
\def\Rl{{\mathbb R}}\def\Cx{{\mathbb C}}
\def\Ir{{\mathbb Z}}\def\Nl{{\mathbb N}}
\def\Rt{{\mathbb Q}}
\def\norm #1{\Vert #1\Vert}
\def\mod{{\mathop{\rm mod}\nolimits}}
\def\ind{{\mathop{\rm ind}\nolimits}\,}
\def\bra #1{\langle #1\vert}
\def\ket #1{\vert #1\rangle}
\def\braket #1#2{\langle #1 \vert #2\rangle}
\def\ketbra #1#2{\vert #1\rangle \langle #2\vert}
\def\kettbra#1{\ketbra{#1}{#1}}
\def\tr{\mathop{\rm Tr}\nolimits}
\def\abs#1{\vert#1\vert}
\def\rank{{\mathop{\rm rank}\nolimits}\,}
\def\ttr{{\mathop{\bf Tr}\nolimits}} 

\mathchardef\ree="023C \mathchardef\imm="023D  

\def\AA{{\mathcal A}}
\def\BB{{\mathcal B}}
\def\CC{{\mathcal C}}
\def\DD{{\mathcal D}}
\def\RR{{\mathcal R}}\def\HH{{\mathcal H}}
\def\NN{{\mathcal N}}\def\KK{{\mathcal K}}
\def\MM{{\mathcal M}}
\def\FF{{\mathcal F}}
\def\TT{{\mathcal T}}
\def\LL{{\mathcal L}}
\def\JJ{{\mathcal J}}

\def\Alf{{\mathbb A}}
\def\fatasterisk{\lower 4pt\hbox {\Large\bf\char42}}
\def\bigtimessym{{\diagup\mkern-17mu\diagdown\mkern2mu}}
\def\bigtimes{\mathop\bigtimessym}

\def\PP{{\mathbf P}} 

\def\ovlp{\eta}

\def\Spp{{\bf S}} 

\newtheorem{thm}{Theorem}
\newtheorem{lem}[thm]{Lemma}
\newtheorem{prop}[thm]{Proposition}
\newtheorem{defn}[thm]{Definition}

\def\sbsection#1{\par\vskip12pt plus24pt\noindent{\it#1\/}.---\ }

\begin{document}

\title{Index theory of one dimensional quantum walks and cellular automata}
\author{D.\ Gross\inst{1}\inst{2}\inst{4} \and
V.\ Nesme\inst{1}\inst{3}\inst{5} \and
H.\ Vogts\inst{1}\inst{6} \and
R.F.\ Werner\inst{1}\inst{7}
}
\institute{
Inst. f. Theoretical Physics, Univ.\ Hannover, Appelstr. 2, 30167 Hannover, Germany \and
Inst.\ f.\ Theoretical Physics, ETH Z\"urich, 
Wolfgang-Pauli-Strasse 27, 8093 Z\"urich, \and
Univ. Potsdam, Inst.\ f.\ Physik, 
K. Liebknecht Str.\ 24/25, 14476 Potsdam-Golm, Germany \and 
www.phys.ethz.ch/\~{ }dagross \and
nesme@qipc.org \and
holger.vogts@gmx.de \and
reinhard.werner@itp.uni-hannover.de
}

\maketitle

\begin{abstract}
	If a one-dimensional quantum lattice system is subject to one step
	of a reversible discrete-time dynamics, it is intuitive
	that as much ``quantum information'' as moves into any given block
	of cells from the left, has to exit that block to the right. For two
	types of such systems --- namely quantum walks and cellular
	automata --- we make this intuition precise by defining an
	\emph{index}, a quantity that measures the ``net flow of quantum
	information'' through the system. The index supplies a complete
	characterization of two properties of the discrete dynamics.  First,
	two systems $S_1, S_2$ can be ``pieced together'', in the sense that
	there is a system $S$ which acts like $S_1$ in one region
	and like $S_2$ in some other region, if and only if $S_1$ and $S_2$
	have the same index. Second, the index labels connected components
	of such systems: equality of the index is necessary and sufficient
	for the existence of a continuous deformation of $S_1$ into $S_2$.
	In the case of quantum walks, the index is integer-valued, whereas
	for cellular automata, it takes values in the group of positive
	rationals. In both cases, the map $S \mapsto \ind S$ is a group
	homomorphism if composition of the discrete dynamics is taken as
	the group law of the quantum systems.  Systems with trivial index
	are precisely those which can be realized by partitioned unitaries,
	and the prototypes of systems with non-trivial index are shifts.
\end{abstract}
\section{Introduction}

Quantum walks and quantum cellular automata are quantum lattice
systems with a discrete step dynamics, which is reversible, and
satisfies a causality constraint: In each step only finitely many
neighboring cells contribute to the state change of a given cell.
This leads to an interesting interplay between the conditions of
reversibility (unitarity) and causality, which is the subject of this
article.

Starting point of the analysis is a simple intuition: for any
connected group of cells in a one dimensional system as much ``quantum
information'' as moves into the subsystem from the left has to move
out at the other end. Moreover, this ``flow'' is a conserved quantity,
in the sense that it remains constant over the spatial extent of the
system. It can thus be determined locally at any point.

Making this intuition precise, we associate with every such lattice
system an \emph{index}, a quantity measuring the net flow of
information.
The index theory developed in this work completely resolves three, a
priori very different, classification problems:

(1) \emph{Find all locally computable invariants}. It is shown that
there exists a ``crossover'' between two systems $S_1, S_2$ if and
only if their indices coincide. More precisely, a crossover $S$
between $S_1$ and $S_2$ is a system which acts like $S_1$ on a
negative half line $\{x|x\leq a\}$ and like $S_2$ on a
positive half line $\{x|x\geq b\}$. Clearly, a locally computable
invariant must assign the same value to two systems if there exists a
crossover between them. It follows that any invariant is
a function of the index.

(2) \emph{Classify dynamics up to composition with local unitaries}.
A natural way of constructing dynamics which respect both
reversibility and causality is by concatenating layers of block
unitaries. In every step, one would decompose the lattice into
non-overlapping finite blocks and implement a unitary operation within every
block. Such \emph{local unitary implementations} are conceptually related to
the gate model of quantum information. Not every time evolution may be
realized this way: a uniform right-shift of cells serves as the
paradigmatic counter-example. We show that the systems with local
implementations are precisely those with trivial index. Consequently,
equivalence classes of dynamics modulo composition with block
unitaries are labeled by their indices.

(3) \emph{Determine the homotopy classes}. It is proved that two
systems may be continuously deformed into each other (with a uniform
bound on the causality properties along the connecting path) if and
only if they have the same index.

We will consider the above questions, and define indices, for two
kinds of systems. {\em Quantum walks} are, on the one hand, the
quantum analogs of classical random walks. On the other hand, they are
discrete time analogs of a standard quantum particle ``hopping'' on a
lattice according to a Hamiltonian which is a lattice version of the
momentum operator $\mathrm{i}\partial_x$. The index defined for these
systems is the same as a
quantity called ``flow'' by Kitaev \cite{Kitaev}. Intuitively, this measures the mean speed of
a quantum walk, expressed in units of ``state space dimensions shifted to the right per time step''.
The mathematical background has been explored, in a more abstract setting, by Avron, Seiler and Simon \cite{Seiler}.
Kitaev's work treated the first classification problem
above. We will re-prove his results with an eye on generalizations to
cellular automata, and will supply solutions to questions (2) and (3).
Although the quantum walks seemed to be comparatively straightforward initially,
the intuition gained from this case
served us well in setting up the theory for the much more involved
case of cellular automata. This allowed us to build an abstract index
theory covering both cases with almost identical arguments
(Sects.~\ref{sec:Gimp} and \ref{sec:Glci}).

{\it Cellular automata} are characterized by the property that
whatever state is possible in one cell (e.g., a superposition of
empty/occupied) can be chosen independently for each cell. Expressed
in terms of particles this means that we are necessarily looking at a
``gas'' system of possibly infinitely many particles. The basic
definition of quantum cellular automata was given in \cite{qca}.  On
the one hand, the setting considered here is more restrictive than
\cite{qca}, covering only one-dimensional systems. On the other hand,
we are allowing for non-translationally invariant dynamics --- a strong
generalization over the earlier paper.  In fact, having completed the
present work we feel that the translation invariance assumed in
\cite{qca} was obscuring the fundamental interplay between
reversibility and causality.  Accordingly, we obtain here a stronger
structure result, even though it is built on the same key ideas.
Throughout, there is a strong interplay between local and global
properties. For example, the following statement is an immediate
consequence of our main Theorem~\ref{thmIndA}: If a nearest neighbor
cellular automaton has somewhere a cell of dimension $n$, and
somewhere else a cell of dimension $m$, coprime to $n$, then it can
be globally implemented as a product of two partitioned unitary
operations.

Our paper is organized as follows: After giving two examples in
Sect.~\ref{sec:ex}, We begin by a mathematical description of what we
mean by quantum walks (Sect.~\ref{sec:Swalks}) and quantum cellular
automata (Sect.~\ref{sec:Sautoms}). We then describe the notion of
locally computable invariants, and why they should form an abelian
group (Sect.~\ref{sec:Glci}). A similar general explanation of the
notion of local implementation is given in Sect.\ref{sec:Gimp}. The
detailed theory for quantum walks is in Sect.~\ref{sec:Iwalk}, and in
Sect.~\ref{sec:Iauto} for cellular automata. This includes the proof
that an index previously defined in the classical translationally
invariant case \cite{Kari_index}, coincides with our index for this
special case (Sect.~\ref{sec:CCA}).  We close with an outlook on
variants of index theory for either higher dimensional systems or
automata with only approximate causality properties
(Sect.~\ref{sec:out}).

\section{Examples}
\label{sec:ex}

Before introducing the mathematical setting, we will illustrate the
problems treated in this paper by giving two concrete examples.

\subsection{Particle hopping on a ring}

The simplest example is given by a single particle on a ring of $N$
sites arranged in a circle. More precisely, the Hilbert space we are
considering is $\mathcal{C}^N$ with basis vectors $\{\ket{e_0}, \dots,
\ket{e_{N-1}}\}$. The vector $\ket{e_i}$ is taken to represent a ``particle
localized at position $i$''. One step of a reversible discrete-time
dynamics is simply given by an arbitrary unitary
$U\in U(\mathcal{C}^N)$. We will consider two such time evolutions
$U_0, U_1$
defined with respect to the standard basis by
\begin{equation}
	U_0: \ket{e_i} \mapsto \ket{e_i},
	\qquad
	U_1: \ket{e_i} \mapsto \ket{e_{(i+1) \,\mathrm{mod}\,N}}
\end{equation}
respectively. The first unitary is the trivial evolution and the
second one models a uniform movement of the particle with velocity one
site per time step to the right.

The physical interpretation of this simple model seems clear: we can
think of a lattice version of a particle with one spatial degree
of freedom, where we have introduced cyclic boundary conditions to get
a simple, finite
description. The
causality property defining a quantum walk then expresses
the physically reasonable assumptions that couplings are local and
dynamics preserve locality.

In this setting, it is natural to think of the time evolution as being
generated by a Hamiltonian: $U_t = e^{\mathrm{i} t H}$. Such a
Hamiltonian formula would allow us to extend the dynamics to arbitrary
real times
$t\in\mathcal{R}$.
To recover $H$, we need to take a logarithm of $U_1$. This operation
is of course not uniquely defined, but the ansatz
\begin{equation}\label{eqn:fourierHopping}
	H=\sum_{k=0}^{N-1} \frac{k}{2\pi} \ket{f_k}\bra{f_k}
\end{equation}
in terms of the Fourier basis
\begin{equation}
	\ket{f_k} = \frac{1}{\sqrt N} \sum_{j=0}^{N-1} e^{\frac{2 \pi
	\mathrm{i}}{N} k j}
\end{equation}
seems particularly appealing. It can easily be checked to be
compatible with our previous definitions of $U_0$ and $U_1$.

\begin{figure}\centering
\includegraphics[width=.18\textwidth]{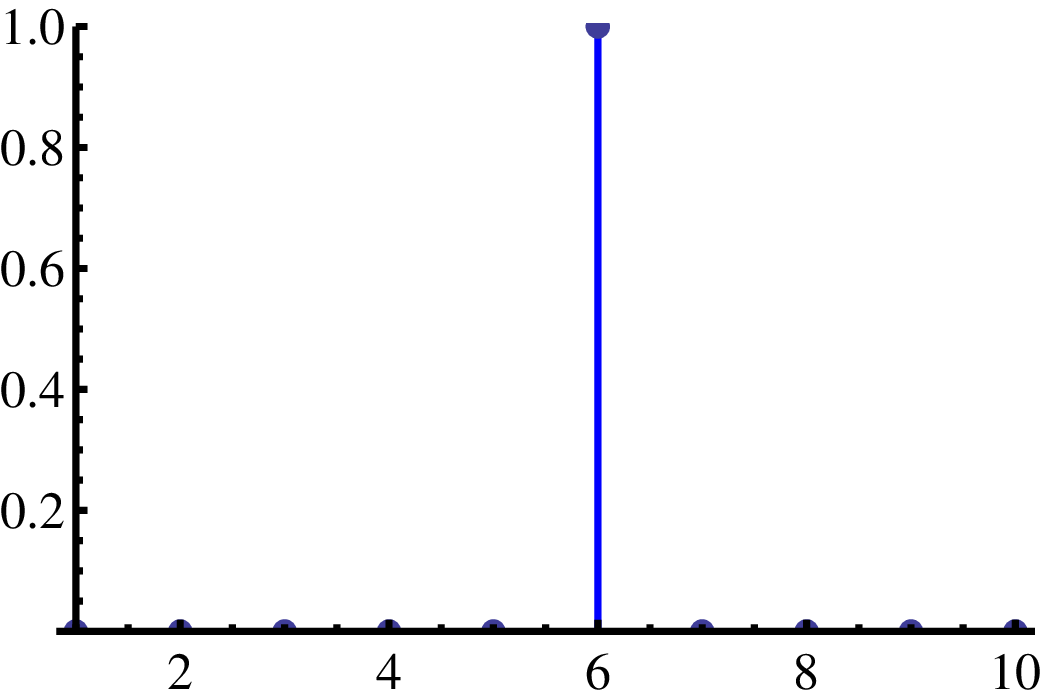}\hspace{.2cm}
\includegraphics[width=.18\textwidth]{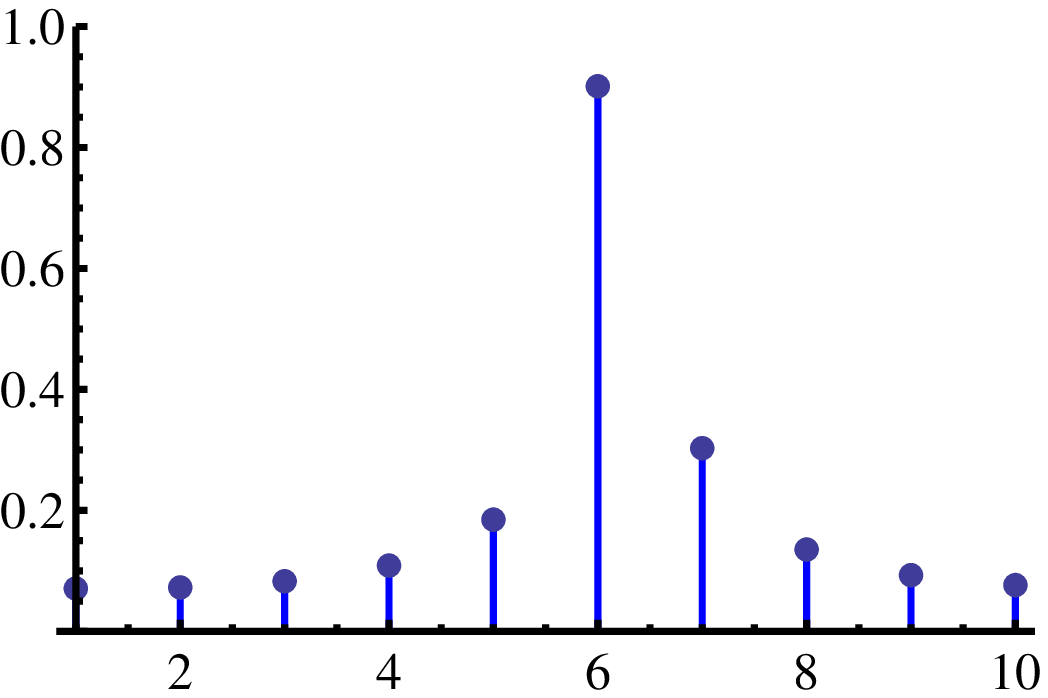}\hspace{.2cm}
\includegraphics[width=.18\textwidth]{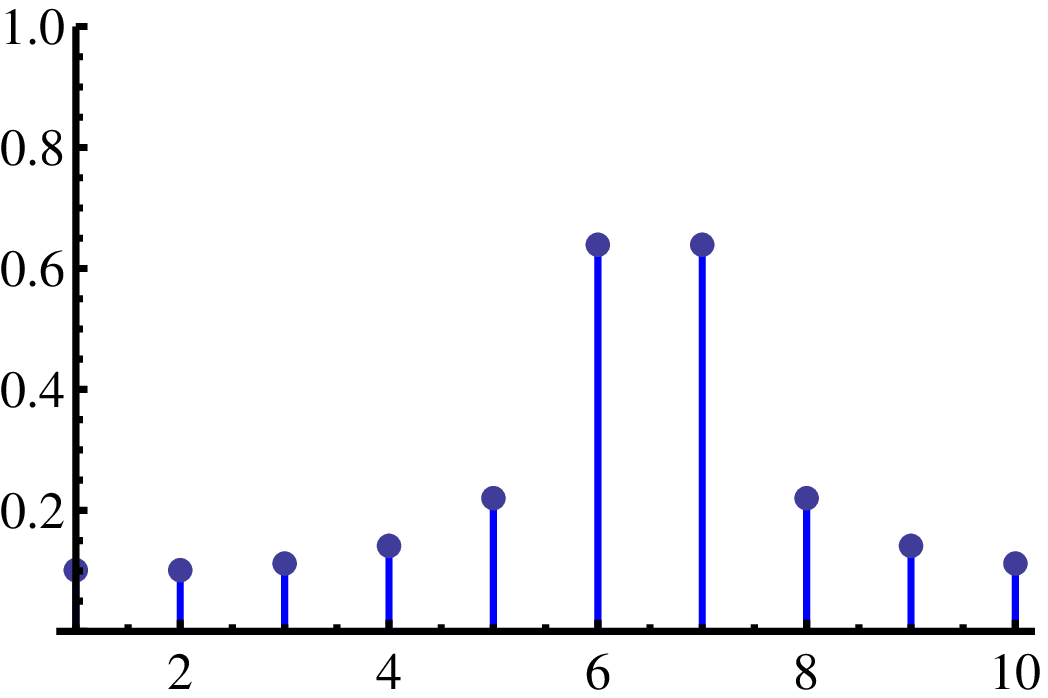}\hspace{.2cm}
\includegraphics[width=.18\textwidth]{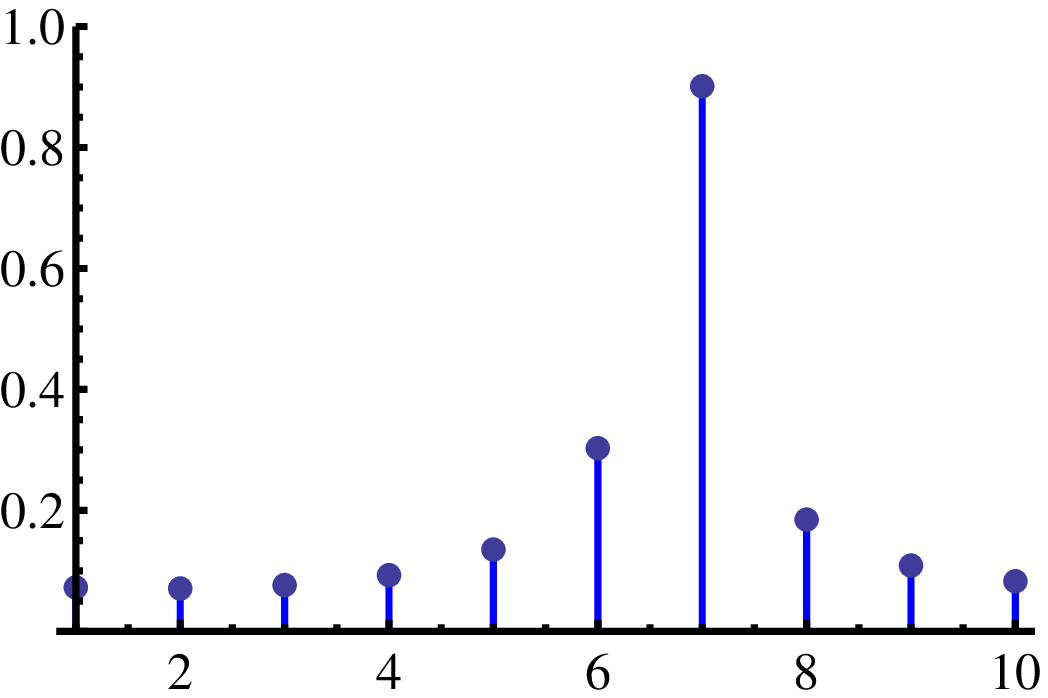}\hspace{.2cm}
\includegraphics[width=.18\textwidth]{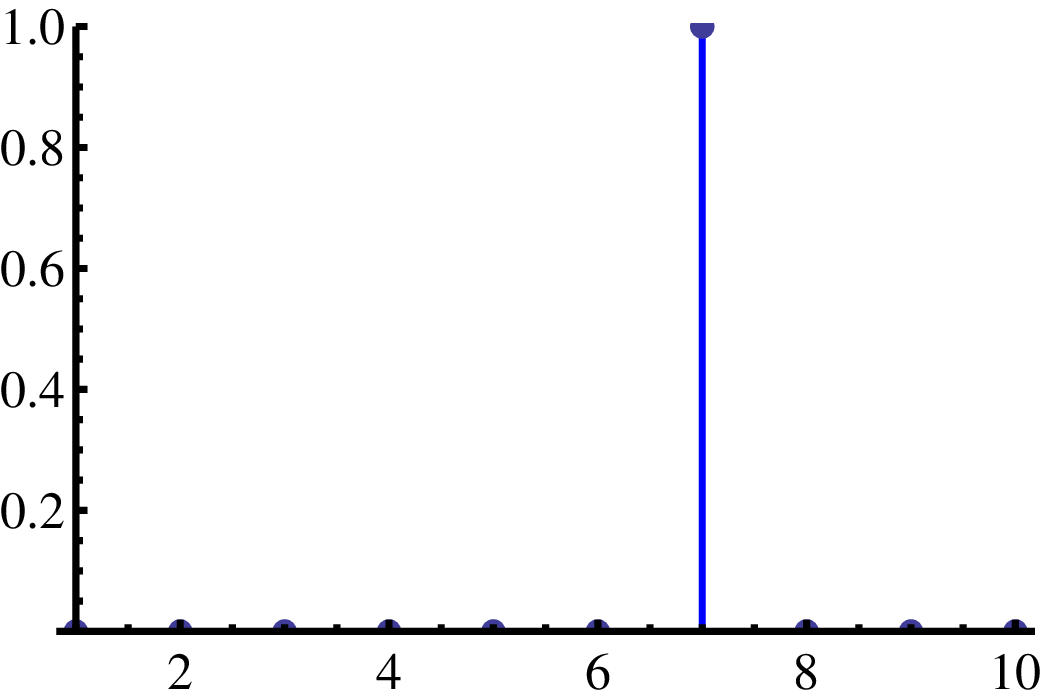}
\caption{Absolute value squared of a vector initially localized at
site $6$ (of $10$) under the action of the time evolution
$e^{\mathrm{i}tH}$ for $t\in\{0,.25,.5.,.75,1\}$. The Hamiltonian is
given by Eq.~(\ref{eqn:fourierHopping}). The dynamics is not causal for non-integer values of $t$.
\label{fig:fourierHopping} }
\end{figure}

Does this definition yield physically satisfactory dynamics for all
times $t\in[0,1]$? Hardly. As demonstrated in
Fig.~\ref{fig:fourierHopping}, a particle initially localized at site
$6$ ($N=10$) will spread out over the entire ring during the interval
$t\in[0,\frac12]$, and refocus to site $7$ during $t\in[\frac12, 1]$.
Any sensible notion of ``causality'' is violated for non-integer values
of $t$. Note that this contrasts with the time evolution generated by
the momentum operator $H=\mathrm{i} \partial_x$ of a 
continuous-variable system. The latter fulfills
$\big(e^{\mathrm{i}t H} \psi\big)(x) =
\psi(x - t)$, thus preserving the localization properties of vectors
$\psi\in L^2(\mathcal{R})$ for any $t\in\mathcal{R}$.

The discussion immediately raises several questions:

Is there a way to continuously interpolate between $U_0$ and
$U_1$ while preserving causality? Our particular choice
(\ref{eqn:fourierHopping}) for $H$ could have been unfortunate.
Conceivably, there is a better-suited, potentially time-dependent
Hamiltonian which does the job. More physically phrased: is it
possible to discretize the spacial degrees of freedom, but not the
temporal ones, of a free single particle,
while respecting causality? (The answer is: no, this is not possible).

Is there a simple way of deciding that the answer to the first
question is negative? (Yes: the index of $U_0$ is $0$, the index of
$U_1$ is $1$. The index is constant on connected components with
uniformly bounded interaction length).

Further questions we will answer include:
Is there a ``gate-model'' implementation of $U_1$? More precisely, can I
write $U_1$ in terms of a constant-depth sequence of unitary
operations, each of which acts non-trivially only on a constant number
of sites? (No).
Can I engineer a large system on $N'\gg N$ sites, endowed with a
global time evolution $U$, such that the restriction of $U$ to some
region of contiguous sites looks like $U_0$ and the restriction of $U$
to some other region looks like $U_1$? (No).

All these statements are made precise in Theorem~\ref{thmInd}.

\subsection{Cluster state preparation}

In this section, we consider $N$ spin-$1/2$ particles arranged on a
circle. The Hilbert space of the system is $\mathcal{H} =
\bigotimes_{i=1}^N \Cx^2$. In the previous example, it was clear what
it meant for a state vector $\psi$ to be ``localized'' in some region
$\Lambda\subset[1,N]$: namely this was the case if $\langle
\psi|e_i\rangle = 0$ for all $i\not\in\Lambda$. For state vectors on
tensor product spaces, on the other hand, there seems to be no
satisfactory notion of ``locality''. To circumvent this problem, we
focus on observables instead. An observable
$A\in\mathcal{B}(\mathcal{H})$ is localized in a region $\Lambda$ if
$A$ acts like the identity on all tensor factors outside of $\Lambda$.

Reversible dynamics on such a system is again represented by a unitary
$U\in U(\mathcal{H})$. We use the symbol $\alpha$ to denote  the
action by conjugation of $U$ on observables: $A \mapsto \alpha(A)= U A U^*$.

Let $\{\sigma_x^{(i)}, \sigma_y^{(i)}, \sigma_z^{(i)}\}$ be the Pauli
matrices acting on the $i$th spin. Since one can form a basis of
$\mathcal{B}(\mathcal{H})$ from products of the Pauli matrices $\{\sigma_x^{(i)},
\sigma_z^{(i)}\}_{i=1}^N$ acting on single spins alone, it suffices to
specify the effect of $\alpha$ on these $2 N$ matrices in order to
completely determine the dynamics. For example, we can set
\begin{eqnarray}
	\alpha(\sigma_x^{(i)}) &=&
	\sigma_z^{(i-1)}\otimes
	\sigma_x^{(i)}\otimes
	\sigma_z^{(i+1)}, \\
	\alpha(\sigma_z^{(i)}) &=& \sigma_z^{(i)}.
\end{eqnarray}
It is a simple exercise to verify that the operators on the right
hand side fulfill the same commutation relations as the $\{\sigma_x^{(i)},
\sigma_z^{(i)}\}$. This is sufficient to ensure that a unitary $U$
implementing the time evolution $\alpha$ actually exists.

As in the previous example, we can ask more refined questions about
$U$. For example: can we implement $U$ by a fixed-depth circuit of
nearest-neighbor unitaries? Can one interpolate between $U$ and the
trivial time evolution while keeping observables localized along the
path? Is there a simple numerical invariant which would allow us to
easily decide these questions?

In this particular case an educated guess gives rise to affirmative
answers to all these questions. Indeed, set
\begin{equation}
	V_t^{(i, i+1)} =
	\left(
		\begin{array}{cccc}
			1 & 0 & 0 & 0 \\
			0 & 1 & 0 & 0 \\
			0 & 0 & 1 & 0 \\
			0 & 0 & 0 & e^{\mathrm{i} t \pi}
		\end{array}
	\right)
\end{equation}
with respect to the standard basis of the $i$th and $(i+1)$th spin.
Define
\begin{equation}
	U_t = \prod_{i=1}^N V_t^{(i,i+1)}
\end{equation}
(the product is well-defined because $V_t$ commutes with its
translates, so the order in which the product is taken is immaterial).
Then one checks that $\alpha$ corresponds to $U_1$, whereas $U_0$ is
the trivial evolution. Clearly $U_t$ continuously interpolates between
these two cases, and, by construction, does not increase the
localization region of an observable by more than two sites.  We note
that $U_1$ is known in the quantum information literature as the
interaction used to generate \emph{graph states}
\cite{SchlingemannWerner,HeinEisertBriegel}.

In Section~\ref{sec:Iauto}, we will set up a general theory for
answering the questions posed above---including in cases where one is
not so lucky to have an explicit parametrization in terms of gates
at hand.

\section{Systems}\label{sec:prelims}%
\subsection{Quantum Walks}\label{sec:Swalks}%
We consider a quantum system with a spatial degree of freedom
$x\in\Ir$, and at every site or ``cell'' $x$ a finite dimensional ``one-cell Hilbert space'' $\HH_x$.
These spaces can be thought of as describing  the internal states of the system as opposed to the
external, spatial variables $x$.
The Hilbert space of the system is
\begin{equation}\label{dirsum}
 \HH=\bigoplus_{x=-\infty}^\infty \HH_x,
\end{equation}
The specification of the $\HH_x$ will be called a {\em cell structure}.

We call a unitary operator $U$ on $\HH$ {\em causal}~{}\footnote{We
are indebted to a referee who drew our attention to the inflationary use of the word ``local'' in our manuscript, where,
among other things, quantum walks were called ``local'' unitaries. We changed this to ``causal'' for the
crucial finite propagation property of walks and cellular automata. This is reminiscent of relativistic propagation
in algebraic quantum field theory and in keeping with usage in signal processing
(where $x$ would be time, and $x_-=x$). In quantum information it agrees with \cite{semilocal,ANW07}.
The terminology disagrees with what some field theorists would probably say \cite{Wer87a}, and with \cite{qca}.}, %
or a {\em quantum walk},
if, for any $x\in\Ir$, there are some $x_-<x_+$ such that
$\phi\in\HH_x\subset\HH$ implies
$U\phi\in\bigoplus_{y=x_-}^{x_+}\HH_x$. We assume that $x_\pm$ are
both non-decreasing as functions of $x$, and go to $\pm\infty$ when
$x$ does.

By $U_{yx}:\HH_x\to\HH_y$ we denote the block matrix corresponding to
the direct sum (\ref{dirsum}), i.e.
\begin{equation}\label{walklocal}
    U\bigoplus_x\phi_x=\bigoplus_y\sum_xU_{yx}\phi_x.
\end{equation}
The causality of $U$ implies that, for any $x$, only finitely many $y$
give non-zero summands.

\sbsection{Grouping}
The spatial variable $x$ of a walk and the internal degrees of
freedom, described in $\HH_x$, are largely interchangeable. In one
direction we can choose a basis $\ket{x,1}, \ldots, \ket{x,d}$ in some
$\HH_x$, and replace the point $x$ by the sequence of points $(x,1),
\ldots, (x,d)$, each with a one-dimensional space $\HH_{x,i}$ of
internal states. Because in the above definition, we assumed that the
spatial variable ranges over $\Ir$, groupings have to be followed up
by a relabeling of sites in the obvious way.
In the other direction, we can ``fuse together'' several cells
$x_1,\ldots,x_k$, getting a new cell $X$ with internal state space
$\HH_X=\bigoplus_{i=1}^k\HH_i$. In either case it is clear how to
adjust the neighborhood parameters $x_\pm$.

Hence we can either regard our system as one without internal degrees of freedom, and Hilbert space $\HH=\ell^2(\Ir)$. Typically this may involve some large neighborhoods $[x_-,x_+]$. Or else, we can group cells until we get a nearest neighborhood system, i.e., $x_\pm=x\pm1$, at the expense of having to deal with high-dimensional $\HH_x$.

Most of these definitions and constructions are easily generalized to
higher-dimensional lattices. It is therefore instructive to identify
the feature which restricts our results to the one-dimensional case.
Indeed, it lies in the fact that one can choose a partitioning
into intervals $[a_i,b_i]\subset\Ir$
such that the sites below $a_i$ and above $b_i$ interact only through
the interval $[a_i,b_i]$. Formally, that is a consequence of demanding
$\lim_{x\to\pm\infty}x_\pm=\pm\infty$.
It is easy to see that such a separation need not be possible in a
two-dimensional lattice, even if every cell has a finite neighborhood.
In this more general setup, neighborhood relations up to regroupings
may be described in terms of \emph{coarse geometry} \cite{coarse}, a
theme we will not pursue here.

\sbsection{Translationally invariant walks}
The simplest way to define a cell structure is to choose a Hilbert
space $\HH_0$, and to set $\HH_x\equiv\HH_0$ for all $x\in\Ir$. We
then have the unitary equivalence $\HH\cong\ell^2(\Ir)\otimes\HH_0$.
In that case we can define the {\em shift} operation $S$ and its
powers by
\begin{equation}\label{shift}
    S^n(\ket x\otimes\phi_0)=\ket{x+n}\otimes\phi_0.
\end{equation}
In this setting one frequently looks at translationally invariant walks, i.e., unitaries $U$ commuting with $S$. More generally, there might be some period $p$ such that $[U,S^p]=0$. Clearly, it is natural in this case to group $p$ consecutive cells, so that after grouping one gets a strictly  translationally invariant walk.

The space $\HH_0$ can then either be considered as an internal degree
of freedom of a walking particle, or as a {\em coin} so that shift
steps (possibly depending on the internal state) are alternated with
unitary coin tosses $\idty\otimes C$. Translationally invariant
systems will be treated in more detail in Sect.~\ref{sec:tiw}.

\sbsection{Periodic boundary conditions}
Since we are after a local theory of quantum walks, global aspects ---
like the distinction between
a walk on $\Ir$ and a walk on a {\it large ring} of $M$ sites ---  are
secondary, as long as the interaction length $L=\max\abs{x_+-x_-}$ remains small in
comparison with $M$. In fact, from any walk on a ring we can
construct one on $\Ir$ which locally looks the same. More formally,
let the sites of the ring be labeled by the classes $\Ir_M$ of
integers modulo $M$, and identified with $\{0,\ldots,M-1\}$. Then we
extend the cell structure by setting $\HH_{x+kM}=\HH_x$ for all
$k\in\Ir$. In order to extend the unitary $U$ on the ring to a walk
$\widehat U$ on $\Ir$ we set
\begin{equation}\label{Uperiodic}
    \widehat U_{xy}=\left\lbrace
       \begin{array}{cl}
          0& \quad \mbox{if }\ |x-y|>L         \\
          U_{x'y'} &\quad \mbox{if }\  |x-y|\leq L,\
                 x'\equiv x, y'\equiv y\ \mod M.
       \end{array}\right.
\end{equation}
For the second line to be unambiguous, we require
$2L<M$. To verify unitarity we need to compute
\begin{equation}\label{UperiodicUnitary}
    \sum_y\widehat U_{xy}^*\widehat U_{yz}
      =\sum_{y'}U_{x'y'}^*U_{y'z'},
\end{equation}
where we have assumed that $|x-z|\leq 2L$, because otherwise the left
hand side is zero anyhow. Note that for each summation index $y$ only
one class $y'\in\Ir_M$ can occur in the sum on the right hand side.
Moreover, every class $y'$ appears, although possibly with a zero
contribution. But the sum on the right is $\delta_{x'z'}\idty_{x'}$,
which together with a similar argument for $UU^*$ proves the unitarity
of $\widehat U$.

From the point of view of index theory, the walk $U$ on the ring
and $\widehat U$ on the line are the same. However, if we iterate $U$, the
interaction length $(x_+-x_-)$ increases, and eventually non-zero matrix elements can
occur anywhere in $U^n$. In this sense, the set of quantum walks on a ring does not form a group.
This is the reason why the theory of walks on $\Ir$ is more elegant and more
complete. From now on we will therefore consider walks on $\Ir$ only.

\subsection{Cellular Automata}\label{sec:Sautoms}%
Once again we consider a system in which a finite dimensional Hilbert
space $\HH_x$ is associated with every site $x\in\Ir$. However,
rather than combining these in a direct sum, we take their tensor
product. In plain English this means that, for any two sites $x,y$,
rather than having a system of type $\HH_x$ at position $x$ {\it or}
a system of type $\HH_y$ at position $y$, as in a quantum walk, we
now have a system of type $\HH_x$ at position $x$ {\it and} a system
of type $\HH_y$ at position $y$. In contrast to the infinite direct
sum of Hilbert spaces, the infinite tensor product is not
well-defined. Since we want to look at local properties, we could
work with a ``potentially'' infinite product, i.e., some finite
product with more factors added as needed in the course of an
argument. But it is easier to work instead with the observable
algebras $\AA_x$, equal to the operators on the Hilbert space
$\HH_x$, or equivalently the algebra $\MM_{d(x)}$, where
$d(x)=\dim\HH_x$ ($\MM_d$ denotes the algebra of $d\times
d$-matrices). In analogy to the definition for walks, we will refer to
the specification of the algebras $\AA_x$  as the
{\em cell structure}.

For the observable algebras associated to sets
$\Lambda\subset\Ir$, we use the following notations: for finite $\Lambda$, $\AA(\Lambda)$
is the tensor product of all $\AA_x$ with $x\in\Lambda$. For
$\Lambda_1\subset\Lambda_2$ we identify $\AA(\Lambda_1)$ with the
subalgebra $\AA(\Lambda_1)\otimes
   \idty^{\Lambda_2\setminus\Lambda_1}\subset\AA(\Lambda_2)$. For
infinite $\Lambda\subset\Ir$ we denote by $\AA(\Lambda)$ the
C*-closure of the increasing family of finite dimensional algebras
$\AA(\Lambda_f)$ for finite $\Lambda_f\subset\Lambda$, also called
the quasi-local algebra \cite{BraRo}. In particular, the algebra of the whole chain is $\AA(\Ir)$, sometimes abbreviated to $\AA$.

A {\em cellular automaton} with cell structure $\{\AA_x\}_{x\in\Ir}$
is an automorphism $\alpha$ of $\AA=\AA(\Ir))$ such that, for some
functions $x\mapsto x_\pm$ as specified in Sect.~\ref{sec:Swalks},
each $\alpha(\AA_x)\subset\AA([x_-,x_+])$. Note
that the restricted homomorphisms $\alpha_x:\AA_x\to\AA([x_-,x_+])$ uniquely
determine $\alpha$, because every observable acting on a finite number of cells can
be expanded into products of one-site observables. These {\em local
rules} $\alpha_x$ have to satisfy the constraint that the algebras
$\alpha_x(\AA_x)$ for different $x$ commute element-wise. In that case
they uniquely determine an endomorphism $\alpha$. For examples and
various construction methods for cellular automata we refer to
\cite{qca}.

Exactly as in the case of quantum walks we can group cells together for
convenience. Whereas the dimensions for subcells add up for
quantum walks
($\dim\bigoplus_{x\in\Lambda}\HH_x=\sum_{x\in\Lambda}\dim\HH_x$) we
get $\AA(\Lambda)\cong\MM_d$ with the product
$d=\prod_{x\in\Lambda}d(x)$.

By considering the time evolution of observables, we have implicitly
chosen to work in the Heisenberg picture.  The expectation value of
the physical procedure ({\it i}) prepare a state $\rho$, ({\it ii})
run the automaton for $k$ time steps, ({\it iii}) measure an
observable $A$ would thus be given by the expression
$\rho(\alpha^k(A))$.  Accordingly, we choose a convention
for the {\em shift} on a chain with isomorphic cells, which at first
seems inverted relative to the definition (\ref{shift}) for walks. We
define it as the automorphism $\sigma$ with $\sigma(\AA_x)=\AA_{x-1}$,
acting according to the assumed isomorphism of all the cell algebras.
Thus if one prepares a certain state, it will be found shifted to the
right after one step of $\sigma$, in accordance with (\ref{shift})
although in that case $U\HH_x=\HH_{x+1}$.

\section{Local Implementability}
\label{sec:Gimp}

We have defined the causality properties of walks and cellular automata
\emph{axiomatically}, i.e., as a  condition on the input-output
behavior of the maps $U$ and $\alpha$. Alternatively, one may take a
\emph{constructive} approach. Here, one would list a set of operations
that should certainly be included in the set of local dynamics, and
refer to any given time evolution as being locally implementable if it can be
represented as sequence of these basic building blocks.  Both methods
are equally valid, and in this section we will completely analyze
their relation.
From the axiomatic point of view this might
be called a ``structure theorem'', 
whereas
from the constructive point of view one would call it a
``characterization theorem''.

In the case at hand, there is a natural choice of building blocks.
Namely, we can partition the system into some subsets (``blocks'') of
sites, and apply a unitary operation separately to each subsystem in
the partition. (Note that the unitaries would be combined by a direct
sum for walks and by a tensor product for cellular automata). For such
maps the interplay between unitarity and causality is trivial: causality
puts no constraint whatsoever on the choice of unitaries acting in
each block. Moreover, it allows the overall operation to be resolved
into a sequence of steps, in which one block operation is done after
the other.  This picture is close to the gate model of quantum
computation~\cite{nc}: here each block unitary would correspond to one
``gate'' involving some subset of registers, so that these gates do
not disturb each other. The fact that they can be executed in parallel
is expressed by saying that these infinitely many gates nevertheless
represent an operation of {\em logical depth 1}.

For partitioned unitary operations the various block unitaries
obviously commute. Commutation is really the essential feature if we want to
resolve the overall time step into a sequence of block unitary steps. Indeed,
consider a family of commuting unitaries $U_j$, each localized in a
finite subset $\Lambda_j$ of some infinite lattice (not necessarily
one-dimensional). We only need that the cover by the $\Lambda_j$ is
locally finite, i.e., each point $x$ is contained in at most finitely
many $\Lambda_j$. Then the product $\prod_j U_j$ implements a
well-defined operation on localized elements. In the cellular
automaton case (where localization just means $U_j\in\AA(\Lambda_j)$),
we define the action on a local observable $A$ as
\begin{equation}\label{communitary} \alpha(A)=\Bigl(\prod_j
U_j^*\Bigr)A \Bigl(\prod_j U_j\Bigr), \end{equation} with the
understanding that both products range over the same index set, namely
those $j$ for which $\Lambda_j$ meets the localization region of $A$.
Here the products can be taken without regard to operator ordering,
since we assumed that the $U_j$ commute. Including additional factors
$U_j$ on both sides will not change $\alpha(A)$, since such factors
can be ``commuted past'' all other $U_{j'},U_{j'}^*$ and $A$ to meet the
corresponding $U_j^*$ and cancel. So the product is over {\it all}
$j$, in the sense of a product over any sufficiently large finite set.
Similar considerations apply for the case of walks on general
lattices.

Now if a QCA $\alpha$ is represented in the form (\ref{communitary}), we can also represent it as a product of partitioned operations: indeed, we only need to group the $U_j$ into families within which all $\Lambda_j$ are disjoint. The product of each family is obviously a partitioned unitary and under suitable uniformity conditions on the cover we only need a finite product of such partitioned operations to represent $\alpha$, typically $s+1$ factors, where $s$ is the spatial dimension of the lattice.
Hence we consider the representation as a product of partitioned unitaries as essentially equivalent to the representation by commuting unitaries as in (\ref{communitary}). In either case we will say that the system is {\em locally implementable}.

We now come to the basic result for implementing general walks or cellular automata by commuting unitaries --- provided we are allowed to enlarge the system. The key feature of these extensions is that they work in arbitrary (not necessarily one-dimensional) lattices and that the ancillary system is a copy of the system itself, on which we implement simultaneously the inverse operation. In the following result, we allow the underlying ``lattice'' $X$ to be any countable set. General neighborhood schemes are described as controlled sets in some coarse structure \cite{coarse}. For the present paper it suffices to describe causality in terms of a metric $d$ on $X$, of which we only assume that all balls  $\NN_L(x)=\{y|d(x,y)\leq L\}$ are finite sets. The causality condition for walks on $X$ is then that there is some ``interaction radius'' $L$ such that in (\ref{walklocal}) $U_{yx}=0$ for $d(x,y)>L$. Similarly, for QCAs, the causality condition reads $\alpha(\AA_x)\subset\AA(\NN_L(x))$. For walks $U,V$ on the same lattice we simply write $U\oplus V$ for a walk with one-cell Hilbert spaces $\HH_x\oplus\KK_x$, where $\HH_x$ are the one-cell spaces for $U$ and $\KK_x$ those for $V$. This splits the total Hilbert space into $\HH\oplus\KK$, and $U\oplus V$ acts according to this direct sum. Similarly, we define the tensor product $\alpha\otimes\beta$ acting on two parallel systems combined in a tensor product.

\begin{prop}\label{ANW}
\hfill\break
(1) For any quantum walk $U$, the walk $U\oplus U^*$ is locally implementable.
\hfill\break
(2) For any cellular automaton $\alpha$, the automaton $\alpha\otimes\alpha^{-1}$ is locally implementable.
\end{prop}

\proof
(1) We are considering a doubled system in which the one-cell Hilbert space at $x$ is $\HH_x\oplus\HH_x$. Let $S_x$ denote the unitary operator on the doubled system which swaps these two summands, and acts as the identity on the one-cell spaces of all other sites. Now consider the unitaries
$$  T_x=(U^*\oplus\idty)S_x(U\oplus\idty). $$
These commute, because they are the images of the commuting transformations $S_x$ under the same unitary conjugation. Moreover, they are localized near $x$ by the causality properties postulated for $U$. Hence their infinite product defines a walk unitary, as discussed above. This unitary is
$$ \prod_xT_x=(U^*\oplus\idty)S(U\oplus\idty)=S(\idty\oplus U^*)(U\oplus\idty)=S(U\oplus U^*), $$
where we have used that $S=\prod_xS_x$ is just the global swap of the two system copies. Hence
$U\oplus U^*=(\prod_xS_x)(\prod_xT_x)$ is locally implemented.

\par\noindent
(2) Essentially the same idea works for cellular automata \cite{ANW07}. Again we consider the unitaries
$S_x\in\AA_x\otimes\AA_x$, which swap the two tensor factors, so that $S_x(A_x\otimes B_x)=(B_x\otimes A_x)S_x$. Now consider the unitary elements
$$ T_x=(\id\otimes \alpha)[S_x].$$
Here we have written the arguments of an automorphism in brackets, to distinguish it from grouping parentheses, and thus eliminate a possible source of confusion in the coming computations.
As images of a family of commuting unitaries under an automorphism,
the $T_x$ are themselves a commuting family of unitaries. Moreover, they
are localized
in $\AA_x\otimes\AA_{\NN(x)}$. Hence they implement a cellular
automaton $\beta$. We determine it by letting it act first on a localized
element of the form $A_x\otimes\idty$ with $A_x\in\AA_x$.
\begin{eqnarray}\nonumber%
 \beta[A_x\otimes\idty]&=&\Bigl(\prod_yT_y\Bigr)^*\bigl(A_x\otimes\idty\bigr)\Bigl(\prod_yT_y\Bigr)\\
                &=&(\id\otimes \alpha)\Bigl[\prod_yS_y\Bigr]\
                 (\id\otimes \alpha)\Bigl[A_x\otimes\idty\Bigr]\
                   (\id\otimes \alpha)\Bigl[\prod_yS_y\Bigr]   \nonumber\\
                &=&(\id\otimes \alpha)\Bigl[(\prod_yS_y)(A_x\otimes\idty)(\prod_yS_y)\Bigr]
                                 \nonumber\\
                &=&(\id\otimes \alpha)[\idty\otimes A_x]=\idty\otimes\alpha[A_x]
                \nonumber
\end{eqnarray}
A similar computation shows that $\beta[\idty\otimes \alpha[B_x]]=B_x\otimes\idty$. Since $\alpha$ is an automorphism, this is the same as $\beta[\idty\otimes B_x]=\alpha^{-1}[B_x]\otimes\idty$. Using the homomorphism property of $\beta$, we get for general localized elements $A,B$ that
$\beta[A\otimes B]=\alpha^{-1}[B]\otimes\alpha[A]$. Hence following $\beta$ by a global swap (implemented by $\prod_xS_x$) we have implemented  $\alpha\otimes\alpha^{-1}$ locally.
\qed

\section{The group of locally computable invariants}
\label{sec:Glci}

In this section we take up the idea of a {\it locally computable
invariant} and show that, for either walks or automata, these
invariants necessarily form an abelian group. The group multiplication reflects both the composition and the parallel application to a double chain. We postpone to later sections the question whether nontrivial invariants exist, i.e., at this stage it might well be that the group
described here is trivial. Later on we will determine this
group to be $(\Ir,+)$ for quantum walks (see
Sect.~\ref{sec:Iwalk})and $(\Rt_+,\cdot)$, the multiplicative group
of positive fractions, for cellular automata (see
Sect.~\ref{sec:Iauto}). In this section, in order not to double each
step, we will describe the arguments for the case of walks, and just
comment at the end on the necessary changes for the cellular automaton case.

Suppose we have defined a property $\PP(U)$, which is defined for
any quantum walk $U$, and which can be determined solely on the basis
of a finite collection of the block matrices $U_{xy}$. More
specifically, if we write the walk in nearest-neighbor form by grouping, we call
a property $\PP(U)$ {\em locally computable} if we can compute it from the restriction of $U$ on any
interval of length $\geq2$. The crucial part of this definition is, of course, that the result obtained in this way must be the same for any interval we may select for the computation, a property which we stress by calling $\PP$ a locally computable invariant.

Suppose now that two walks $U_1$ and $U_2$ {\it share a patch}, in
the sense that there is a long interval $[x_1,x_2]\subset\Ir$, on which the Hilbert spaces $\HH_x$ for $x\in[x_1,x_2]$ have the same dimensions and, after the choice of a suitable isomorphism,
the unitaries $U_1$ and $U_2$ restricted to these subspaces act in the same way. We assume that the interval is sufficiently long to determine $\PP$. Then
local computability just means that we must have $\PP(U_1)=\PP(U_2)$. In
other words, $\PP$ must be constant on each equivalence class of the relation of
``sharing a patch''. So the theory of locally computable invariants is really
equivalent to characterizing the classes of the transitive hull of
this relation:  We will write $U_1\sim U_n$, if there is a chain of
walks $U_1,U_2,\ldots,U_n$ such that, for all $i$, $U_i$ and
$U_{i+1}$ share a patch. In contrast to the relation of sharing a
patch, this equivalence relation no longer makes any requirements about the sizes of any one-cell Hilbert spaces in the walks $U_1$ and $U_n$.
The most comprehensive locally computable property is now just the property
of $U$ to belong to some equivalence class: all other locally computable properties are functions of this class. Our aim thus shifts to computing the set $\JJ$ of equivalence classes for  ``$\sim$''. The equivalence class of a walk $U$ will be denoted by $\ind(U)\in\JJ$, and  called its (abstract) {\em index}.

\sbsection{Triviality of locally implementable systems}
Let us first make the connection to the questions of the previous section: suppose that a walk or automaton is locally implementable, i.e., the product of a collection of block partitioned unitaries.
Compare this with a system in which all unitaries, whose localization intersects the positive half axis, are replaced by the identity. Clearly, this acts like the identity on all cells on the positive axis, and we can further modify the system by making it trivial ($0$-dimensional $\HH_x$ or $1$-dimensional algebras $\AA_x$) for $x>0$. Clearly these systems share a large patch (most of the negative axis), so they are equivalent. In other words, locally implementable systems have the same index as the identity on a trivial chain.

\sbsection{Crossovers}
A very useful fact about the relation
$U_1\sim U_2$ is that it implies a prima facie much stronger relation:
It is equivalent to the property that there is a ``crossover'' walk $U_c$, which coincides with $U_1$ on a negative half line $\{x|x\leq a\}$ and coincides with $U_2$ on a positive half line $\{x|x\geq b\}$.

\proof
Let us denote the relation just described by $U_1\approx U_2$. Then
$U_1\approx U_2\Rightarrow U_1\sim U_2$, because $U_1$ and $U_2$ each share an infinite patch with $U_c$.

In the converse direction, if $U_1$ and $U_2$ share a patch, we can define $U_c$ to be the walk whose one-cell Hilbert spaces $\HH_x$ are those of $U_1$ for $x$ to the left of the shared patch,
and are those of $U_2$ for $x$ to the right of the shared patch.
Similarly, we define the unitary $U_c$ to coincide with $U_1$ to the
left and with $U_2$ on the right. On the shared patch we can choose
either one, since the two walks coincide. Since the shared patch was
assumed to be sufficiently long this does not lead to an
ambiguity for either $U_c$ or $U_c^{-1}$. Hence $U_1\approx U_2$.

In order to cover the case that $U_1$ and $U_2$ are linked by a chain in which any neighbors share a patch, we only need to prove that $U_1\approx U_2$ is a transitive relation.

\begin{figure}\centering
\includegraphics[width=8.5cm]{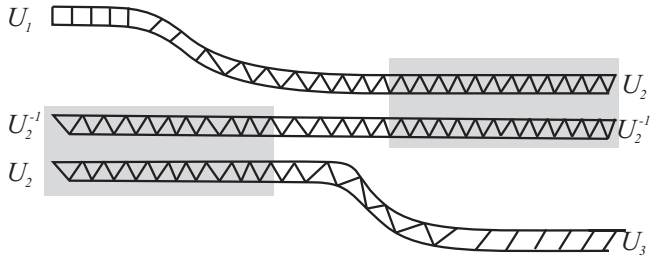}
\caption{Combining a crossover from $U_1$ to $U_2$ with a crossover from $U_2$ to $U_3$ to obtain a crossover from $U_1$ to $U_3$. The shaded double chains can be fused to a single cell by Prop.~\ref{ANW}.}
\label{fig:crosstrans}
\end{figure}

In order to prove transitivity, consider the walks $U_1,U_2,U_3$ with crossovers $U_{12}$ and $U_{23}$ as shown in Fig.~\ref{fig:crosstrans}. There we also included a copy of the inverse of $U_2$. We take the overall picture as a representation of the walk $U_{12}\oplus U_2^{-1}\oplus U_{23}$. Now consider the strands of $U_{12}$ and $U_2^{-1}$ to the right of the crossover region of $U_{12}$. Since $U_2\oplus U_2^{-1}$ is locally implementable by Prop.~\ref{ANW}, we can replace this pair of strands by a trivial system, still retaining a legitimate unitary operation for the rest. Similarly, we can fuse the strands of $U_{23}$ and $U_2^{-1}$ to the left of the crossover region of $U_{23}$ to nothing. This results in a unitary operator, which coincides with $U_1$ on a left half axis and with $U_3$ on a positive half axis, i.e., a crossover $U_{13}$.
\qed

\sbsection{Grouping} Suppose we regroup some finite collection of the
cells. Clearly, this does not affect cells far away, and we
immediately get a crossover. Hence the index does not change when regrouping cells, even if this is carried out in parallel. We will
implicitly use this in the sequel by regrouping sites in whatever way is most convenient.

\sbsection{Parallel chains}
Define the direct sum $U\oplus V$ of walks as in the previous section. Then if $U_c$ (resp.\ $V_c$) is a crossover between $U_1$ and $U_2$ (resp.\ $V_1$ and $V_2$), $U_c\oplus V_c$ is obviously a crossover between $U_1\oplus V_1$ and $U_2\oplus V_2$. Hence the class $\ind(U\oplus V)\in\JJ$ depends only on the equivalence classes of $U$ and $V$, and we can define an ``addition'' of indices by $\ind(U)+\ind(V)=\ind(U\oplus V)$.
This addition is abelian, because there is a trivial crossover between $U\oplus V and V\oplus U$, just exchanging the summands on a half chain.  Moreover, since an inverse is defined as $-\ind(U)=\ind(U^*)$ via Prop.~\ref{ANW}, we conclude that $\JJ$ becomes an abelian group.

\sbsection{Products}
Now suppose $U$ and $V$ are walks on the same cell structure, so that $UV$ makes sense. We claim that there is a crossover between $UV$ and $U\oplus V$. Hence we also get
$$\ind(UV)=\ind(U\oplus V)=\ind(U)+\ind(V).$$

Indeed, consider the cell structure on which $U\oplus V$ is defined, which has one-cell Hilbert spaces $\HH_x\oplus\HH_x$ at each site. Now let $S_+$ denote the unitary which acts as the swap on all $\HH_x\oplus\HH_x$ with $x>0$ and leaves the subspaces with $x\leq0$ unchanged. Then $(U\oplus\idty)S_+(\idty\oplus V)S_+$ is a crossover between $U\oplus V$ and $UV\oplus\idty\sim UV$.

\sbsection{Modifications for cellular automata}
The concept of crossovers and the arguments for the group structure
can be taken over verbatim,
with the replacements $U\mapsto\alpha$, $\HH_x\mapsto\AA_x$, $\oplus\mapsto\otimes$.
Of course, the index group will also be different, and we will adopt the convention to write it multiplicatively. The product formula thus reads $\ind(\alpha\beta)=\ind(\alpha\otimes\beta)=\ind(\alpha)\ind(\beta)$.

\sbsection{Numerical representation and shift subgroups}
To make the abstract theory of this section useful, one needs to establish an isomorphism of the index group with some explicitly known group. The natural way to do that is to identify {\it generators}, i.e., some particular walks which cannot be implemented locally, but are sufficient to generate arbitrary walks up to locally implementable factors. Although it is far from obvious at this point, it will turn out later that, for walks as well as for cellular automata, the role of generators is played by the shifts. Since there is only one kind of generators, it suffices to assign numbers as ``index values'' to the shifts to establish an isomorphism of the abstract index groups with groups of numbers.

For walks, the shift was introduced in
Sect.~\ref{sec:Swalks}. We denote by $S_d$ the shift on a system with
$d$-dimensional internal degree of freedom ($\dim\HH_x=d$ for all
$x$). Similarly, let $\sigma_d$ denote the shift automorphism on a
chain with cell algebra $\MM_d$.
We tentatively demand
\begin{equation}\label{shiftindW}
    \ind(S_d)=d
\end{equation}
and, similarly,
\begin{equation}\label{shiftindA}
    \ind(\sigma_d)=d.
\end{equation}
This has to be consistent for shifts on parallel chains. Since $S_d\oplus S_e=S_{d+e}$ and
$\sigma_d\otimes\sigma_e=\sigma_{d\cdot e}$, this requires that we take the indices of walks as a group of numbers under addition, and for the cellular automata as a group of numbers under multiplication.
Indeed, we will show that the above formulas fix an isomorphism of the abstract index group $\JJ$ to
$(\Ir,+)$ for quantum walks, and to the group $(\Rt_+,\cdot)$ of positive fractions for cellular automata.


\section{Index for quantum walks}
\label{sec:Iwalk}

\subsection{Pedestrian definition} The following is the basic
definition of this chapter.  To the best knowledge of the authors, it
is due to Kitaev \cite{Kitaev}, who calls this quantity the {\em flow}
of a walk $U$.

\begin{defn}\label{defInd}
For any walk $U$, we define the {\bf index} as
\begin{equation}\label{defIndW}
   \ind U=\sum_{x\geq0>y}\left(\tr (U_{xy})^*U_{xy}-\tr (U_{yx})^*U_{yx}\right).
\end{equation}
\end{defn}
Note that the sum is finite by virtue of the definition of causal
unitaries. Clearly, for the simple shift we get $\ind  S_1=1$,
confirming Eq.~(\ref{shiftindW}).

Of course, we will show presently that this quantity has all the properties of the abstract index discussed in the previous sections. However, from the definition given here it seems miraculous that
such a quantity should be always an integer, and independent of the
positioning of the cut. To see this it is better to rewrite this
quantity in the following way.

\subsection{Operator theoretic definition}
We introduce the projection $P$ for the half axis $\{x\geq0\}$, i.e.,
the projection onto the subspace $\bigoplus_{x\geq0}\HH_x$. Then, for
$\phi_x\in\HH_x$ and $\phi_y\in\HH_y$ we get
\begin{equation}
   \bra{\phi_x} PU-UP\ket{\phi_y}=\left\lbrace
     \begin{array}{cl}
        0       & x\geq0\ \mbox{and}\ y\geq0\\
        -\bra{\phi_x}U_{xy}\ket{\phi_y}  & x<0\ \mbox{and}\ y\geq0\\
        \bra{\phi_x}U_{xy}\ket{\phi_y} & x\geq0\ \mbox{and}\ y<0\\
        0       & x<0\ \mbox{and}\ y<0
     \end{array}\right.
\end{equation}
Hence, for any pair $(x,y)$ the commutator $[U,P]$ has just the
signs used in the definition of the index, and we get
\begin{equation}\label{indcomm}
   \ind U=\tr U^*[P,U]=\tr(U^*PU-P)
\end{equation}
Note that for the the trace on the right hand side we cannot use
linearity of the trace to write it as the difference of two (equal!)
terms, because this would result in an indeterminate expression
$\infty-\infty$.

\subsection{Fundamental properties of the index for walks}

\begin{thm}\label{thmInd}\
\begin{enumerate}
\item $\ind U$ is an integer for any walk $U$
\item $\ind U$ is locally computable, and uniquely
    parameterizes the equivalence classes for the relation $\sim$
    from Sect.~\ref{sec:Glci}, hence can be identified with the abstract
    index defined there.
\item $\ind(U_1\oplus U_2)=\ind(U_1)+\ind(U_2)$, and, when $U_1$ and $U_2$ are defined on the same cell structure, $\ind(U_1U_2)=\ind(U_1)+\ind(U_2)$.
     Moreover, for the shift of $d$-dimensional cells: $\ind S_d=d$.
\item $\ind U=0$ if and only if $U$ admits a ``local
    decoupling'', i.e. there is a unitary $V$, which acts like the
    identity on all but finitely many $\HH_x$, such that $UV$ is block diagonal with respect to the
    decomposition $\HH=\bigl(\bigoplus_{x\leq0}\HH_x\bigr)\oplus\bigl(\bigoplus_{x\geq1}\HH_x\bigr)$.
\item $\ind U=0$ if and only if it is locally implementable (see Sect.~\ref{sec:Gimp}).
    In this case it can be written as a product of just two partitioned unitaries. When $U$ is regrouped in nearest neighbor form, then the partitioned unitaries can be chosen to couple only pairs of nearest neighbors.
\item $\ind U_1=\ind U_0$ if and only if $U_0$ and $U_1$ lie in the same
    connected component, i.e.,  there is a norm continuous path
    $[0,1]\ni t\mapsto U_t$ of causal unitaries of uniformly
    bounded interaction length $L$ with the specified boundary
    values.
\end{enumerate}
\end{thm}

\noindent The rest of this subsection is devoted to the proof of this
result. According to Eq.~(\ref{indcomm}), the index is closely related
to a difference of projections. If these were finite dimensional, we
could just use linearity to get the difference of two integers.
The following Lemma shows that the result is still an integer when
the difference of the two projections has finite rank. Actually it
is even sufficient for the difference to be trace class, and with a
careful discussion of the trace, it is sufficient for the $\pm1$ eigenspaces
of the difference to be finite dimensional \cite{Seiler}. Here we include
the simple case sufficient for our purposes.

\begin{lem}\label{finrank}
Let $Q,P$ be orthogonal projections in a Hilbert space $\HH$, such
that $Q-P$ has finite rank. Then
\begin{enumerate}
\item The range $\RR=(Q-P)\HH$ is an invariant subspace for both
    $Q$ and $P$.
\item  $\tr(Q-P)$ is an integer.
\item There is a unitary operator $V$ such that $V\phi=\phi$ for
    all $\phi\perp(Q-P)\HH$, and such that $Q\geq V^*PV$ or
    $Q\leq V^*PV$.
\item If $\tr(Q-P)=0$, the $V$ from the previous item satisfies
    $Q=V^*PV$.
\end{enumerate}
\end{lem}

\proof
1. follows from the identity $$Q(Q-P)=Q(\idty-P)=(Q-P)(\idty-P),$$
and its analogue for $P$. \\
2. Clearly, we can evaluate the trace in a basis of $\RR$ since the
basis elements from $\RR^\perp$ contribute only zeros. Since the
restrictions of $Q$ and $P$ to $\RR$ are projections on a finite
dimensional space,
$$\tr(Q-P)=\tr_\RR(Q-P)=\tr_\RR(Q)-\tr_\RR(P)$$
is the difference of two natural numbers.\\

3\&4. Obviously, we can find such a unitary on $\RR$ with the
corresponding property for the restrictions of $Q$ and $P$ to $\RR$.
We then extend $V$ to be the identity on $\RR^\perp$. When
$\tr(Q-P)=0$, this $V$ is a unitary mapping from $Q\RR$ to $P\RR$.
\qed

\proof[of Theorem~\ref{thmInd}]
None of the statements, or values of the index will change under grouping, except part of item 5, which requires nearest neighbor form. Therefore we will assume without loss that all walks are nearest neighbor.
We will use Lemma~\ref{finrank} with $Q=U^*PU$.\\

1.  This follows directly from Eq.~(\ref{indcomm}) and
Lemma~\ref{finrank}, item 2.

2. Let $P'$ be the projection onto another half axis, say $x\geq x_0$.
Then $P-P'$ is finite rank and hence
$(U^*PU-P)-(U^*P'U-P')=P'-P-U^*(P-P')U$ is the difference of two
finite rank  operators with equal trace. Hence the index does not
depend on the cut position, and since formula~(\ref{defInd}) clearly
involves only matrix elements at most 1 site from the cut, it is a
locally computable invariant. It remains to be shown that it is a
complete invariant, i.e., that $\ind U_1=\ind U_2$ implies $U_1\sim
U_2$ in the sense of Sect.~\ref{sec:Glci}. This will be done in
connection with item 4 below.

3. This follows from Sect.~\ref{sec:Glci}. But a direct proof
(for the product) is also instructive:
$$(U_1U_2)^*P(U_1U_2)-P=(U_2^*PU_2-P)+U_2^*(U_1^*PU_1-P)U_2$$
is the sum of two finite rank operators, of which we can take the
trace separately.

4. Apply Lemma~\ref{finrank}, item 3, to get $V$ with $P=V^*(U^*PU)V$, and
hence $PUV=UVP$. The fact that $V-\idty$ vanishes on all but finitely many
$\HH_x$ follows from its construction: $V-\idty$ vanishes on the complement of
$(P-Q)\HH\subset\HH_{-1}\oplus\HH_0$, for the cut ``$-1|0$'' used in Def.~\ref{defInd}.
Note that this implies
$V\sim\idty$, and also $UV\sim\idty$, since a unitary which has no
matrix elements connecting $x\geq0$ and $x<0$ clearly allows a
crossover with the identity. From the product formula for locally
computable invariants we therefore get that $\ind U=0$ implies
$U\sim\idty$. Obviously, this extends to other values of the index:
if $\ind U_1=\ind U_2$ we have $\ind U_1^*U_2=0$,  hence
$U_1^*U_2\sim\idty$ and hence $U_1\sim U_2$. This completes the proof
of item 2.\\

5\&6. These items each contain a trivial direction: We have already shown in Sect.~\ref{sec:Glci} that locally implementable walks have trivial index. Moreover, it is clear from
Definition~\ref{defInd} that the index is a continuous function, and must hence be constant on each connected component.
The non-trivial statement in 5.\ is that walks with trivial index are indeed implementable, and in 6.\ that walks with vanishing index can be connected to the identity (the rest then follows by multiplication).
In either case, an explicit construction is required, and it will actually be the same one.

So let $U=U_1$ be a walk $\ind U=0$. Let $V_0$ denote the decoupling unitary for the cut $-1|0$, obtained in the proof of item 4, and
define similar unitaries $V_k$ for the cuts at $2k-1|2k$ such that $UV_k$
has no non-zero matrix elements $(UV_k)_{xy}$ with $y<2k\leq x$. Let
$H_k$ denote a hermitian operator located on the same subspaces as
$V_k-\idty$, such that $V_k=\exp(iH_k)$. We will take all $H_k$
bounded in norm by the same constant ($\pi$ will do). Then since they
live on orthogonal subspaces, their sum $H=\sum_kH_k$ is well-defined
and also bounded. Now let $V(t)=\exp(itH)$, which is a norm continuous
function of $t$, because $\norm H\leq\pi$. The endpoint $V(1)$ can
also be defined by this product formula $V(1)=\prod_kV_k$, because on
each subspace $\HH_{2k-1}\oplus\HH_{2k}$ only one these unitaries
is different from $\idty$. Moreover, $UV(1)$ has no matrix elements
$y<2k\leq x$ for any $k$, i.e., it is block diagonal for a
decomposition of $\Ir$ into pairs $\{2k,2k+1\}$. Now take a
similar Hamiltonian path deforming each block in this matrix
decomposition separately to the identity. Specifically, we take $W(0)=\idty$ and
$W(1)=UV(1)$. Then $t\mapsto W(t)V(t)^*$ is a norm continuous path
(although no longer a unitary group), with the endpoints $\idty$ and
$U$. Moreover, each unitary $W(t)$ or $V(t)^*$ is based on a partition into neighboring pairs so that,
for no $t$, $W(t)V(t)^*$ has any non-zero matrix element between sites with $|x-y|>2$.
This proves the remaining statement in item 5 (for $t=1$), and also the statement about uniformly bounded neighborhoods in item 6.
\qed

\subsection{The translation invariant case}
\label{sec:tiw}%
Suppose that $U$ commutes with some power of the shift. It is then
useful to group spaces $\HH_x$ into larger blocks to get commutation
with the shift itself. That is, in this section we assume all
$\HH_x\equiv\KK$ to be equal, and $U_{xy}=U_{x-y}$, where by a
slight abuse of notation the single-index quantity $U_x$ is defined
as $U_{x0}$. The width $L$ is the largest $x$ such that $U_x\neq0$
or $U_{-x}\neq0$. It is natural to diagonalize $U$ using the Fourier
transform. We define
$\FF:\ell^2(\Ir)\otimes\KK\to\LL^2([-\pi,\pi])\otimes\KK$ by
$\FF(\Psi)(p)=\frac1{\sqrt{2\pi}}\sum_xe^{ipx}\Psi(x)$. This is to
be read as a $\KK$-valued equation, where we use the natural identification
of $\LL^2([-\pi,\pi])\otimes\KK$ with the set of $\KK$-valued square
integrable functions on $[-\pi,\pi]$. Similarly, we identify $\ell^2(\Ir)\otimes\KK$
with the $\KK$-valued square summable sequences.
 Then $\FF U\FF^*$ becomes the
multiplication operator by the $p$-dependent matrix
\begin{equation}\label{Uhat}
    \widehat U(p)=\sum_{x=-L}^L U_xe^{ipx}.
\end{equation}
Note that this is a Laurent polynomial in $e^{ip}$. The
largest degree of $e^{ip}$ in the polynomial is $x_+-x$,
which is constant by
translation invariance. The lowest degree is
$x-x_-$. Further, $\widehat U(p)$
must
be a unitary operator on $\KK$ for every $p\in\Rl$.
Taking these facts together, we conclude that both
the
determinant $\det\widehat U(p)=f(p)$
and its inverse
$1/f(p)=\det\widehat U(p)^*$
are Laurent polynomials as well.
But this is only possible if $f$ is actually a
monomial, say proportional to $\exp(inp)$ for some integer $n$. We
claim that this $n$ is the index:

\begin{prop} For a translation invariant walk,
\begin{equation}\label{tiindexDet}
    \det\widehat U(p)=C e^{ip\ \ind(U)},
\end{equation}
for some phase constant $C$.
\end{prop}

As a simple example consider the shift on a chain with $\dim\HH_0=1$. We already noted after Definition~\ref{defInd} that this has index $1$. The corresponding $p$-dependent unitary is the number $\widehat U(p)=e^{ip}$, so this also gives index $1$. For unitaries acting on each site separately in the same way, we get agreement because $\ind U=0$, and $\widehat U(p)$ is independent of $p$.
Note also that both sides of \ref{tiindexDet} have the same behavior under composition and direct sums.
This proves the formula for all walks which can be composed of shifts and sitewise rotations.
Actually, {\it all} translationally invariant walks  can be represented in this way \cite{Gao}, but we prefer to give a direct proof of the proposition without invoking this decomposition.

\proof
From Definition~\ref{defInd} we get
\begin{eqnarray}\nonumber
 \ind U
   &=& \sum_{x=0}^\infty\sum_{y=-\infty}^{-1}\tr|U_{x-y}|^2-\tr|U_{y-x}|^2
     \\\nonumber
   &=& \sum_{n=-\infty}^\infty n\tr|U_n|^2
    =  \sum_{nm}\delta_{nm} n\tr(U_m^* U_n)\\\nonumber
   &=&\frac1{2\pi i}\int_{-\pi}^\pi\!\!dp\ \tr \Bigl(\widehat U(p)^*
             \frac{d\widehat U(p)}{dp}\Bigr)\nonumber
\end{eqnarray}
On the other hand, for any invertible matrix function $\widehat U$,
$$\frac d{dp}\det\widehat U(p)
  =\det\widehat U(p)\ \tr\Bigl(\widehat U(p)^{-1}\frac{d\widehat U(p)}{dp}\Bigr).$$
Hence with $\det\widehat U(p)=\exp(ipn)$ the above integrand is actually constant equal to
$in$, and $\ind(U)=n$.
\qed

\begin{figure}\centering
\psfrag{W}{$\omega_k(p)$} \psfrag{p}{$p$}
\includegraphics[width=8cm]{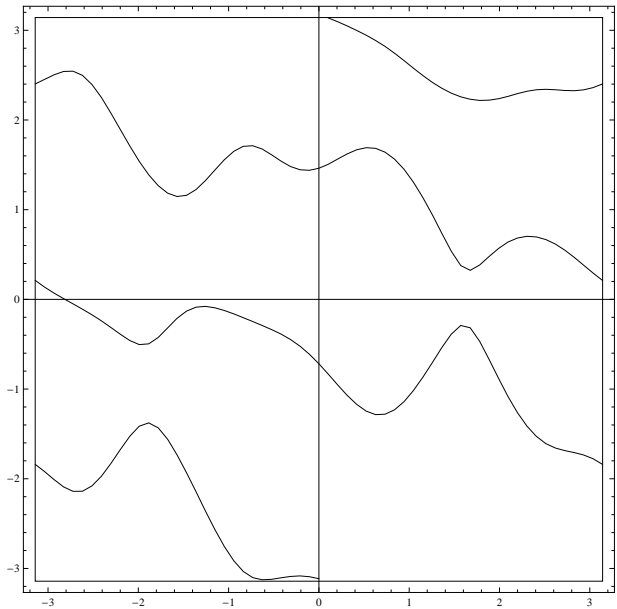}
\caption{\label{fig:wind}Example for eigenvalues of $\widehat U(p)$, with
$\dim\KK=3$, $L=5$, $\ind U=-1$. In this case the spectrum is a
single curve on the torus. The index can be computed by the signed
number of crossings of any horizontal line, or as the sum
of the derivatives of all branches.}
\end{figure}

The properties of a walk depend crucially on the properties of the eigenvalues
$e^{i\omega_1(p)},\ldots,e^{i\omega_d(p)}$ of $\widehat U(p)$ (see Fig.~\ref{fig:wind}). Clearly, $p\mapsto\widehat U(p)$ is an analytic family of operators, so we can follow the eigenvalues as analytic curves \cite{Kato}. The derivatives of the eigenvalues determine the {\em group velocity}: Let $P_\alpha(p)$ be the eigenprojection of $\widehat U(p)$ belonging to the eigenvalue $\exp{i\omega_\alpha(p)}$. Then the group velocity operator can be written as
\begin{equation}\label{groupVelo}
    G(p)=\lim_{t\to\infty}\frac1tX(t)
     =\sum_\alpha P_\alpha(p)\frac{d \omega_\alpha(p)}{d p},
\end{equation}
where $X$ denotes the position operator, and at degenerate eigenvalues the projections $P_\alpha(p)$ have to be chosen appropriately, as dictated by analytic perturbation theory. The limit is in the strong resolvent sense. Hence the probability distribution for the selfadjoint operator $G$ in a state $\rho$ is equal to the asymptotic position distribution starting from $\rho$ in ``ballistic'' scaling\cite{TimeRandom}. In particular, when the internal state is unpolarised, i.e., when the initial state is of the form $\rho=\sigma\otimes\idty/d$,
we find \cite{Kitaev}
$$ \langle X(t)\rangle=\langle X(0)\rangle+ \frac td \ind(U).$$
In this sense the index has direct relevance as a mean speed of the walk.

It is, of course, suggestive to connect the topological properties of the eigenvalue curves in Fig.~\ref{fig:wind} with the index. In principle, these curves are allowed to cross each other. So in general, we have several branches of curves, which wind several times around the torus before closing. The number of curves and their winding numbers would appear to be a topological invariant, but this is not true because of the ``avoided crossing'' phenomenon, in which a small perturbation of the walk turns an eigenvalue crossing into a close encounter of two separate curves
(suggested also at $p=1.6$ in Fig.~\ref{fig:wind}). Hence only the {\em sum of the winding numbers} is stable with respect to small perturbations, and this is indeed equal to $\ind U$. To see this, note that at every $p$ the sum of the derivatives of all branches is equal to the index. Therefore the sum of the winding angles of the branches taken from $p=-\pi$ to $p=\pi$ is $2\pi\,\ind U$. The winding angles of the closed curve components are just suitable sub-sums of this, and this partition is changed at avoided crossing points.

We close this section by establishing a variant of the main theorem for the translationally invariant case. Of course, most of this follows by simply specializing. The only question which requires a new argument is whether the path connecting two translationally invariant walks can be chosen to go entirely through translationally invariant walks. This is established in the following proposition.

\begin{prop} Let $U$ be a translationally invariant causal unitary with $\ind U=0$.
Then we can find a norm continuous path $t\mapsto U_t$ of
translationally invariant causal unitaries of bounded width such that
$U_0=\idty$ and $U_1=U$.
\end{prop}

\proof We use the factorization theorem for ``paraunitary''
operators \cite{Gao} (see also \cite{Vogts}), which states that
$\widehat U(p)$ can be written as a finite product
$$ \widehat U(p)=V_0\prod_{k=1}^r \widehat W_{m_k}(p)\, V_k$$
of constant unitaries $V_k$ and the elementary causal unitaries
$$ \widehat W_m(p)=\left(\begin{array}{cccc}
           e^{imp}&0&\cdots&0\\
           0&1&\cdots&0\\
           \vdots&&\ddots&\vdots\\
           0&0&\cdots&1\end{array}\right)
$$
As a quantum walk these corresponds to a shift of only the first
internal state by $m$ positions, leaving all other internal states
fixed. This walk has index $m$, and the product formula yields $\ind
U=\sum_km_k$. The maximal polynomial degree of any matrix element in
$e^{\pm ip}$ depends on the localization region $[x_+,-x_-]$, and is clearly bounded by $L_{\rm
max}=\sum_k|m_k|$. Hence we can contract the walk to $\idty$ by
contracting each of the $V_k$ to $\idty$, never exceeding width
$L_{\rm max}$ on the way.
\qed

\section{The index for cellular automata}
\label{sec:Iauto}

For cellular automata we proceed in analogy to the case of walks,
i.e., by defining directly a locally computable quantity as the index
$\ind\alpha$ of a walk automorphism $\alpha$. We then establish that
it is actually a complete locally computable invariant and, at the
same time that it characterizes the connected components of QCAs.

As a technical preparation we need some background on commutation
properties of algebras spanning several cells. It is basically
taken from \cite{qca}, and included here to make the presentation
here self-contained.

\subsection{Support algebras} For defining the index we need to find
a quantitative characterization of ``how much'' of one cell ends up
in another. To this end we introduce the notion of support algebras.
Consider a subalgebra $\AA\subset\BB_1\otimes\BB_2$ of a tensor
product. What is the position of $\AA$ relative to the tensor
structure? Here we answer a relatively simple part of this question:
which elements of $\BB_1,\BB_2$ are actually needed to build $\AA$?

For the following Definition with Lemma, recall that $\AA'$ denotes
the commutant $\{a|\forall a_1\in\AA: [a,a_1]=0\}$.
\begin{lem}\label{lem:spp} Let $\BB_1$ and $\BB_2$ be finite dimensional
C*-algebras, and $\AA\subset\BB_1\otimes\BB_2$ a subalgebra. Then
\begin{enumerate}
\item There is a smallest C*-subalgebra $\CC_1\subset\BB_1$ such
    that $\AA\subset\CC_1\otimes\BB_2$. We call this the {\bf
    support algebra} of $\AA$ on $\BB_1$, and denote it by
    $\CC_1=\Spp(\AA,\BB_1)$.
\item Consider a basis $\{e_\mu\}\subset\BB_2$, so that every
    $a\in\AA$ has a unique expansion $a=\sum_\mu a_\mu\otimes
    e_\mu$ with $a_\mu\in\BB_1$. Then $\Spp(\AA,\BB_1)$ is
    generated by all the elements $a_\mu$ arising in this way.
\item The commutant of $\Spp(\AA,\BB_1)$ in $\BB_1$ is
    characterized as $\{b\in\BB_1|b\otimes\idty\in\AA'\}$.
\end{enumerate}
\end{lem}

\proof
We can pick out the terms $a_\mu$ by applying a suitable functional
$\omega_\mu$ from the dual basis to the second factor, i.e., by
applying the map $\id\otimes\omega_\mu:\BB_1\otimes\BB_2\to\BB_1$,
which takes $b_1\otimes b_2$ to $\omega_\mu(b_2)b_1$. Clearly, if
$a\in\CC_1\otimes\BB_2$, so that $a$ can be expanded into simple
tensors using only elements from $\CC_1$ in the first factor, we find
$a_\mu=(id\otimes\omega_\mu)(a)\in\CC_1$. Hence the algebra described
in item 2 must be contained in any $\CC_1$ satisfying item 1. Since
it also satisfies condition 1, we have identified the unique smallest
$\CC_1$. The characterization 3 follows by looking at commutators of
the form $[b\otimes\idty,a]$, and expanding $a$ as above.
\qed

This construction was introduced to the QI community by
Zanardi \cite{Zanardi}, who applied it to the algebra generated by an
interaction Hamiltonian, and consequently called it an ``interaction
algebra''. Of course, we can apply the construction also to the
second factor, so that
\begin{equation}\label{a2Supp}
  \AA\subset \Spp(\AA,\BB_1)\otimes \Spp(\AA,\BB_2)
      \subset\BB_1\otimes\BB_2\;.
\end{equation}

The crucial fact we need about support algebras is that ``commutation
of algebras with overlapping localization happens on the
intersection''. More precisely, we have the following

\begin{lem} \label{sppcomm}
Let $\AA_1\subset\BB_1\otimes\BB_2$ and
$\AA_2\subset\BB_2\otimes\BB_3$ be subalgebras such that
$\AA_1\otimes\idty_3$ and $\idty_1\otimes\AA_2$ commute in
$\BB_1\otimes\BB_2\otimes\BB_3$. Then $\Spp(\AA_1,\BB_2)$ and
$\Spp(\AA_2,\BB_2)$ commute in $\BB_2$.
\end{lem}

\proof Pick bases $\{e_\mu\}\subset\BB_1$ and
$\{e'_\nu\}\subset\BB_2$, and let $a\in\AA_1$ and $a'\in\AA_2$. Then
we may expand uniquely: $a=\sum_\mu e_\mu\otimes a_\mu$ and
$a'=\sum_\nu a'_\nu\otimes e'_\nu$. Then by assumption
\begin{displaymath}
 0=[a\otimes\idty_3,\idty_1\otimes a']
  =\sum_{\mu\nu}e_\mu\otimes [a_\mu,a'_\nu]\otimes e'_\nu \;.
\end{displaymath}
Now since the elements $e_\mu\otimes e'_\nu$ are a basis of
$\BB_1\otimes\BB_3$, this expansion is unique, so we must have
$[a_\mu,a'_\nu]=0$ for all $\mu,\nu$. Clearly, this property also
transfers to the algebras generated by the $a_\mu$ and $a'_\nu$,
i.e., to the support algebras introduced in the Lemma.
\qed

\subsection{Defining the Index}
\label{sec:Defindalpha}

Now consider a cellular automaton $\alpha$ on a cell structure
$\AA_x$. By regrouping, if necessary, we may assume that it has only nearest neighbor interactions.
Now
consider any two neighboring cells $\AA_{2x}\otimes\AA_{2x+1}$, and
their image under $\alpha$, i.e.,
$$ \alpha\Bigl(\AA_{2x}\otimes\AA_{2x+1}\Bigr)\subset
     \Bigl(\AA_{2x-1}\otimes\AA_{2x}\Bigr)\otimes
     \Bigl(\AA_{2x+1}\otimes\AA_{2x+2}\Bigr).$$
We apply the support algebra construction to this inclusion,
obtaining two algebras
\begin{eqnarray}\label{RR2x}
   \RR_{2x}&=& \Spp\Bigl(\alpha\bigl(\AA_{2x}\otimes\AA_{2x+1}\bigr),\
                        \bigl(\AA_{2x-1}\otimes\AA_{2x}\bigr)\Bigr)\\
    \RR_{2x+1}&=& \Spp\Bigl(\alpha\bigl(\AA_{2x}\otimes\AA_{2x+1}\bigr),\
                        \bigl(\AA_{2x+1}\otimes\AA_{2x+2}\bigr)\Bigr)
\end{eqnarray}
Intuitively, the algebras $\RR_y$ with even index become larger when information flows to the right, whereas the ones with odd index describe a flow to the left. Of course, this intuition will be made precise below, but at this stage one can at least check these statements for the shift: When $\alpha(\AA_y)=\AA_{y-1}$ (recall the convention made at the end of Sect.~\ref{sec:Sautoms}) we have $\RR_{2x}=\AA_{2x-1}\otimes\AA_{2x}$ and
$\RR_{2x}=\Cx\idty$. This is to be contrasted with $\RR_y=\AA_y$ for the identity.

Continuing with our construction, observe that by Lemma~\ref{sppcomm}, the subalgebras $\RR_{2x+1}$ and
$\RR_{2x+2}$ commute in the algebra
$\bigl(\AA_{2x+1}\otimes\AA_{2x+2}\bigr)$. Algebras $\RR_y$ which are
further away commute anyhow, since they are contained in disjoint
cells. We conclude that {\em all} $\RR_y$ commute.

By definition of support algebras,
$\alpha(\AA_{2x}\otimes\AA_{2x+1})\subset \RR_{2x}\otimes\RR_{2x+1}$, so that the algebras
$\RR_x$ together generate an algebra containing $\alpha\AA(\Ir)$. Because $\alpha$ is an automorphism, this is the same as $\AA(\Ir)$. Now if
any $\RR_x$ had a non-trivial center, i.e., if there were an element $X\in\RR_x$ commuting with all of $\RR_x$ but not a multiple of $\idty$, this $X$ would also be in the center of the entire quasi-local algebra $\AA(\Ir)$. However, this center is known to be trivial \cite{BraRo}.
We conclude that each
$\RR_x$ must have trivial center, and hence be isomorphic to
$\MM_{r(x)}$ for some integer $r(x)$. Moreover, the inclusion noted at the beginning of this paragraph cannot be strict, since otherwise we would find an element in the relative commutant, which would once again be in the center $\alpha\AA(\Ir)$. To summarize, we must have
\begin{eqnarray}\label{AARReven}
    \alpha\bigl(\AA_{2x}\otimes\AA_{2x+1}\bigr)
     &=&\RR_{2x}\otimes\RR_{2x+1}, \\\mbox{hence}\qquad
     d(2x)d(2x+1)&=&r(2x)r(2x+1).
\end{eqnarray}

\begin{figure}\centering
\psfrag{a}{\Large$\alpha$}
 \psfrag{i0}{$\AA_0$}\psfrag{r0}{$\RR_0$}
 \psfrag{i1}{$\AA_1$}\psfrag{r1}{$\RR_1$}
 \psfrag{i2}{$\AA_2$}\psfrag{r2}{$\RR_2$}
 \psfrag{i3}{$\AA_3$}\psfrag{r3}{$\RR_3$}
 \psfrag{i4}{$\AA_4$}\psfrag{r4}{$\RR_4$}
 \psfrag{i5}{$\AA_5$}\psfrag{r5}{$\RR_5$}
 \psfrag{i6}{$\AA_6$}\psfrag{r6}{$\RR_6$}
\includegraphics[width=8.5cm]{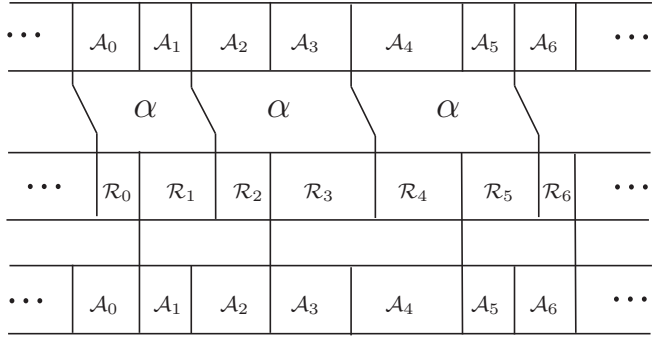}
\caption{\label{fig:algebradims}Cell structure with support algebras. (Read top to bottom)
If the width of cells is taken as log dimension, the index can be
read off the slant in the boxes representing mapping by $\alpha$. }
\end{figure}

On the other hand, the commuting full matrix algebras $\RR_{2x+1}$ and $\RR_{2x+2}$ together span
the tensor product isomorphic to $\MM_{r(2x+1)r(2x+2)}$ inside
$\bigl(\AA_{2x+1}\otimes\AA_{2x+2}\bigr)$. Again the inclusion cannot
be strict, because otherwise the automorphism would not be onto. From
this we get the second relation and dimension equation
\begin{eqnarray}\label{AARRodd}
    \RR_{2x+1}\otimes\RR_{2x+2}
     &=&\AA_{2x+1}\otimes\AA_{2x+2}, \\\mbox{hence}\qquad
     r(2x+1)r(2x+2)&=&d(2x+1)d(2x+2).\nonumber
\end{eqnarray}
These relations are summarized pictorially in Fig.~\ref{fig:algebradims}. They give us the the first two equalities in
\begin{equation}\label{dimtransfer}
    \frac{r(2x)}{d(2x)}=\frac{d(2x+1)}{r(2x+1)}=\frac{r(2x+2)}{d(2x+2)}
    \equiv \ind\,\alpha.\nonumber
\end{equation}
In other words, this quantity is constant along the chain, and hence
we are entitled to define it as the index $\ind \alpha$. The even/odd
asymmetry only comes from the construction, by which the even
$\RR_{2x}$ describe a flow to the right (increasing $x$), and the odd
$\RR_{2x+1}$ are associated with a flow to the left. By shifting the
entire construction, we could switch the even/odd distinction, and
define, for any $\AA_x$, both the ascending and the descending
$\RR_x$. In any case, for the shift $\sigma_d$ of a $d$-dimensional chain we get $d(y)\equiv d$,
$r(2x)=d^2$, and $r(2x-1)=1$, and hence $\ind\sigma_d=d$, as announced in Eq.~(\ref{shiftindA}).

It is part of the local computability property that we have a lot of
freedom in choosing the cell structure for which we want to evaluate
the index. Since one typically wants to use this freedom to simplify
the computation, we will now summarize the constraints. It is clear
that there are three subalgebras involved in the computation, playing
the r\^ole of, say the above $\AA_0,\AA_1,\AA_2$ for determining
$\RR_1$. Let us call these $\AA_L,\AA_M,\AA_R$ to emphasize that
these algebras need not be part of the original cell structure, and
we are free to choose them within certain limits. Let us start by
fixing some algebra $\AA_M\cong\MM_d$, a full matrix algebra
contained in some local algebra, whose crucial property is to split
the system: we must have an isomorphism of the total algebra with
$\AA_{<M}\otimes\AA_M\otimes\AA_{>M}$, where the outer factors
contain suitable infinite half chain algebras, and such that
$$[\alpha(\AA_{<M}),\AA_{>M}]=[\alpha(\AA_{>M}),\AA_{<M}]=\{0\}.$$
Clearly, this imposes a a lower bound on the size of $\AA_M$ in terms
of the interaction length of the automorphism. Now we choose finite
dimensional matrix subalgebras $\AA_R\subset\AA_{>M}$ and
$\AA_L\subset\AA_{<M}$ such that
$\alpha(\AA_M)\subset\AA_L\otimes\AA_M\otimes\AA_R$. These three can
be taken as part of a nearest neighbor cell structure, so that the
above arguments give
\begin{equation}\label{defIndA}
 \left.\begin{array}{rcl}
      \AA_M&\cong&\MM_d\\
      \Spp(\alpha(\AA_L\otimes\AA_M),\AA_M\otimes\AA_R)&\cong&\MM_r
 \end{array}\right\rbrace
 \quad\Rightarrow \ind\alpha=\frac dr
\end{equation}
Note that there is no harm in choosing
$\AA_R$ larger than necessary: the support algebra, being the minimal
algebra needed to build the tensor product, will simply not change.
By a similar argument, we can choose $\AA_L$ too large without
changing this support algebra.

\subsection{Fundamental properties of the index for cellular automata}

\begin{thm}\label{thmIndA}\
\begin{enumerate}
\item $\ind\alpha$ is a positive rational for every $\alpha$.
    When the automaton is regrouped in nearest neighbor form,
    both the numerator and the denominator of $\ind\alpha$ in canceled form
    divide every cell dimension.
\item $\ind\alpha$ is locally computable, and uniquely
    characterizes the equivalence classes for the relation $\sim$
    from Sect.~\ref{sec:Glci}. It can hence be identified with the abstract
    index defined there.
\item $\ind(\alpha\otimes\alpha')=\ind(\alpha)\ind(\alpha')$ and, when $\alpha$ and $\alpha'$ are defined on the same cell structure,  $\ind(\alpha\alpha')=\ind(\alpha)\ind(\alpha')$.\\
    If, for some $y$, $\alpha(\AA((-\infty,y]))\subset\AA((-\infty,y]))$, then $\ind\alpha\in\Nl$. \\
    Moreover, for the shift of $d$-dimensional cells: $\ind \sigma_d=d$.
\item An automaton $\alpha$ has index 1 if and only if it can be implemented locally.
    In this case it can be written as a product of just two partitioned unitary automorphisms. If $\alpha$ is partitioned in nearest neighbor form, the partitioned  automorphisms can be taken to couple pairs of nearest neighbors only.
\item Two automata $\alpha_0,\alpha_1$ on the same cell structure
    have the same index if and only if they  can be deformed to
    each other, i.e., there is a strongly continuous path $[0,1]\ni
    t\mapsto \alpha_t$ of automorphisms, all with the same neighborhoods, and
    with the specified boundary values.
\end{enumerate}
\end{thm}

\proof
(1,3)\ $\ind\alpha\in\Rt_+$  follows immediately from the construction in Sect.~\ref{sec:Defindalpha}, particularly Eq.~(\ref{dimtransfer}). Let $\ind\alpha=\frac pq$ be the fraction in canceled form. Then, from this equation $p=nr(2x)$ and $q=nd(2x)$, where $n$ is the canceled factor. Hence $q$ divides $d(2x)$ and from the second fraction in (\ref{dimtransfer}) we find that $p$ divides $d(2x+1)$. By shifting the construction by one cell, we find the remaining divisibility statements.
From Eq.~(\ref{dimtransfer}) we also get the product formulas 3.
Suppose that $\alpha$ maps some left half chain in to itself. Then we choose a cell partition so that $y=2x+1$ in the setting of (\ref{RR2x}). Then $\RR_{2x+1}\subset\AA_{2x+1}$ and since these are full matrix algebras with the same unit, the quotient $\ind\alpha=d_{2x+1}/r_{2x+1}$ in (\ref{dimtransfer}) is integer. The value for the shift was verified as an example after (\ref{dimtransfer}).

(2,4,5) From Eq.~(\ref{dimtransfer}) local computability is obvious. From the general discussion in Sect.~\ref{sec:Glci} we also get that locally implementable $\alpha$ have $\ind\alpha=1$. The hard part, which is needed to identify the abstract index with the concrete formula is the converse. For this the crucial step is the following claim:

{\em Let $\alpha$ and $\alpha'$ be nearest neighbor cellular automata on the same cell structure and with the same index. Then there are unitaries $V_x\in\AA_x\otimes\AA_{x+1}$ such that the two locally implemented automorphisms
\begin{eqnarray}
    \beta(A)&=&(\prod_xV_{2x-1})^*A(\prod_xV_{2x-1})   \nonumber\\
    \gamma(A)&=&(\prod_xV_{2x})^*A(\prod_xV_{2x})      \nonumber
\end{eqnarray}
satisfy: $\alpha'\gamma=\beta\alpha$}.

Before proving this claim, let us see how it implies the statements in the Theorem. By the general theory of Sect.~\ref{sec:Glci} locally implementable automorphisms $\beta,\gamma$ are $\sim$-equivalent to the identity, and $\alpha\sim\beta\alpha=\alpha'\gamma\sim\alpha'$. Hence equality of the indices as defined by (\ref{defIndA}) implies crossover equivalence, and hence the equality of all locally computable invariants.
This proves item 2. The converse in item 4 follows by taking $\alpha'=\id$, giving the local implementation $\alpha=\beta^{-1}\gamma$ of any index 1 automorphism $\alpha$.

Finally, it is clear for item 5 that we can connect $\alpha$ and
$\alpha'$ with the same index by the required continuous path: we just
need to contract each unitary $V_x$ in $\beta,\gamma$ to the identity,
to obtain a path $\alpha_t=\beta_t\alpha\gamma_t^{-1}$ with
$\alpha_0=\alpha$ and $\alpha_1=\alpha'$. This path will not be
continuous in the norm on automorphisms, i.e., we cannot make
$\norm{\alpha_t-\alpha_s}$ small, since this would already fail for
one-site operations
$\alpha_t(A)=\Bigl(U_t^{\otimes\infty}\Bigr)^*AU_t^{\otimes\infty}$
with $t\mapsto U_t$ norm continuous.  However, for any finitely localized observable $A$, $t\mapsto \alpha_t(A)$ will be continuous in norm, which
is the claim of strong continuity made in the Theorem.  We remark that
an important part of the proof of item 5 is missing at this point:
We did not exclude
the possibility that there are continuous paths linking automorphisms
of different
index.  This will be achieved by Prop.~\ref{davidsformula}, an
expression for the index which is manifestly continuous with respect
to strongly continuous deformations.

Now to prove the claim, let $\alpha$ and $\alpha'$ have the same cell structure and the same index. Then we carry out the construction of Sect.~\ref{sec:Defindalpha} for both automorphisms, resulting in some intermediate algebras $\RR_x$ and $\RR_x'$.

Since the indices coincide, formula (\ref{dimtransfer}) demands that these are full matrix algebras of the same dimensions. For example,
$$ \RR_{2x-1}\otimes\RR_{2x}=\AA_{2x-1}\otimes\AA_{2x}=\RR_{2x-1}'\otimes\RR_{2x}'.$$
Clearly, there is a unitary operator $V_{2x-1}\in\AA_{2x-1}\otimes\AA_{2x}$ so that
$V_{2x-1}^*\RR_{y}V_{2x-1}=\RR'_{y}$ for $y=2x$ and for $y=2x-1$. We can take all these unitaries together as implementing one partitioned automorphism $\beta(A)=(\prod_xV_{2x-1})^*A(\prod_xV_{2x-1})$. By definition, it satisfies $\beta(\RR_y)=\RR'_y$ for all $y\in\Ir$.

Now consider the action of $\alpha$ and $\alpha'$ on
$\AA_{2x}\otimes\AA_{2x+1}$. We now get two isomorphisms
\begin{eqnarray}\nonumber
   \alpha':\AA_{2x}\otimes\AA_{2x+1}&\to&\RR'_{2x}\otimes\RR'_{2x+1} \quad\mbox{and}\\
   \beta\alpha:\AA_{2x}\otimes\AA_{2x+1}&\to&\RR_{2x}\otimes\RR_{2x+1} \to\RR'_{2x}\otimes\RR'_{2x+1}.
   \nonumber
\end{eqnarray}
Hence $(\alpha')^{-1}\beta\alpha$ restricts to an automorphism of $\AA_{2x}\otimes\AA_{2x+1}$, and can therefore be implemented by a unitary $V_{2x}\in\AA_{2x}\otimes\AA_{2x+1}$. These unitaries together implement $\gamma$, and we get the desired equation  $\alpha'\gamma=\beta\alpha$.
\qed

\subsection{Index for classical reversible automata}
\label{sec:CCA}

In this section, we will review a common notion of index for reversible classical cellular automata, and show that it coincides with our definition. In the context of this paper, a reversible classical cellular automaton can be defined as a particular case of a quantum cellular automaton.
In each cell $\AA_x$ we single out a maximal abelian subalgebra
$\DD_x$. With respect to a suitable choice of basis, $\DD_x$ is then
the set of diagonal matrices. As a finite dimensional abelian algebra,
we can regard $\DD_x$ as the set of complex valued functions on a
finite set $\Alf_x$, called the {\it alphabet} of the cell, which at
the same time serves as the set of basis labels for the orthonormal
basis in which $\DD_x$ is diagonal. The global C*-algebra of the
classical system is then the infinite tensor product
$\DD(\Ir)=\bigotimes_{x=-\infty}^\infty\DD_x$. It is canonically
isomorphic to the algebra of continuous functions on the compact
cartesian product space $\Alf_\Ir=\bigtimes_{x=-\infty}^\infty\Alf_x$,
also known as the space of infinite configurations. We use lower case letters such as $c$ for such configurations, and denote by $c(x)\in\Alf_x$ the configuration of the cell at $x$.

Now let $\alpha$ be a QCA with the property that $\alpha(\DD)\subset\DD$. Then the restriction of $\alpha$  to $\DD$ is an automorphism of $\DD$, which must be of the form $(\alpha f)(c)=f(\Phi(c))$, where $\Phi$ is a homeomorphism on configurations. The causality conditions on $\alpha$ are readily expressed in terms of $\Phi$, and, together with analogous arguments for the inverses show that $\Phi$ is a reversible classical cellular automaton in the usual sense, apart from the requirement of translation invariance.
There are some subtle points to note about the correspondence $\alpha\to\Phi$:

\begin{itemize}
\item Suppose that in the above argument we start from a general, not necessarily strictly causal automorphism $\alpha$ of the quasi-local algebra $\AA(\Ir)$. We still get a continuous $\Phi$
    on the compact space $\Alf_\Ir$. By the definition of the product topology this means that the local configurations $\Phi(c)_x$ after the time step depend on only finitely many $c(y)$. In the translation invariant case this means that $\Phi$ is a cellular automaton with finite neighborhood.
    In fact, this argument is used to establish that the inverse of a reversibly cellular automaton also has a finite neighborhood. This is a rather surprising sharpening of the causality condition. However, we are appealing here to a highly non-constructive compactness argument, which gives no control on the size of the neighborhoods, or (barring translation invariance) on the uniformity of the neighborhoods. For example, we can apply it to a cellular automaton $\alpha$ on a 2D lattice, whose quasi-local C*-algebra is isomorphic to that of a 1D automaton. Hence the condition of $\alpha$ being an automorphism is not strong enough to give a 1D automaton in the sense defined above.

\item The mapping is onto, i.e., every classical reversible cellular
automaton can be ``quantized''. The argument is very simple for finite
lattices, e.g., a regular lattice with periodic boundary conditions:
one labels the basis of a Hilbert space by the classical configurations. Then the classical automaton $\Phi$ is a permutation of the basis vectors, which can be interpreted as a unitary operator via $U_\Phi\ket c=\ket{\Phi(c)}$. Then for all observables we set $\alpha(A)=U_\Phi^*AU_\Phi$. One needs to check that this transformation is causal in the quantum sense \cite{qca}, in particular that off-diagonal local operators that are finitely localized (i.e. localized on a finite number of cells) keep this property under the action of $\alpha$.  Indeed one gets a bound on the quantum neighborhoods, which involves both the neighborhoods of $\Phi$ and the neighborhoods of $\Phi^{-1}$. The same computation provides a formula for $\alpha(A)$, for $A$ finitely localized, in terms of the classical rule, and this can be used to define $\alpha$ also for infinite lattices.  We will not, however, make this explicit here.

\item The mapping $\alpha\to\Phi$ is not injective. Indeed an ambiguity is inherent in the construction just described: we can choose different bases with the same diagonal operators $\kettbra a$ by choosing a phase for each basis vector.  This amounts to changing $\alpha$ by an on-site unitary, which is certainly irrelevant for index purposes. But we can consider this more generally: suppose that two cellular automata $\alpha,\beta$ restrict to the same automorphism on the diagonal algebra $\DD$. Then $\alpha\beta^{-1}$ leaves $\DD$ point-wise fixed. In a finite lattice, so that $\alpha$ is unitarily implemented, the implementing unitary hence commutes with all elements of the maximally abelian algebra $\DD$, hence must itself be diagonal. It is suggestive that this also holds in a localized form on the infinite lattice, so $\alpha\beta^{-1}$ would be a product of commuting unitaries and hence have trivial index. The following result builds on this intuition.
\end{itemize}

\begin{prop}\label{classicalfactor}%
Suppose that $\alpha$ and $\beta$ be quantum cellular automata taking the diagonal algebra $\DD$ into itself, with the same restriction to $\DD$.
Then $\ind\alpha=\ind\beta$.
\end{prop}

Before coming to the proof of this proposition we single out two arguments of independent interest, each of which can be used to draw the main conclusion, without discussing in detail the structure of local phase factors. The first criterion uses the absence of propagation.
The second  uses the global transpose map $\Theta:\AA(\Ir)\to\Ir$. It is defined as the matrix transpose on each local algebra, in a basis in which $\DD$ is diagonal. Since transposition is isometric, and consistent with the embeddings $A\mapsto A\otimes\idty$ it extends to the whole algebra. $\Theta$ is a linear anti-homomorphism (meaning $\Theta(AB)=\Theta(B)\Theta(A)$), and, for every automorphism $\alpha$, $\Theta\alpha\Theta$ is again an automorphism.

\begin{lem}
\begin{enumerate}
\item Let $\alpha$ be a nearest neighbor cellular automaton such that, for some finite interval $[z_-,z_+]$ we have $\alpha^n\bigl(\AA([0,1])\bigr)\subset\AA([z_-,z_+])$ for all $n\in\Ir$. Then $\ind\alpha=1$.
\item For any cellular automaton $\alpha$: $\ind(\Theta\alpha\Theta)=\ind\alpha$.
\end{enumerate}
\end{lem}

\proof
1. The index  $\ind\alpha^n$ can be expressed as a ratio of of subcell dimensions of $\AA([z_-,z_+])$, hence is uniformly bounded in $n$.
But since $\ind\alpha^n=(\ind\alpha)^n$ this implies $\ind\alpha\leq1$. With the same argument for the inverse we get $\ind\alpha\geq1$.

2. By assumption, the global transposition is made with respect to  product basis, so that for a tensor product $\BB=\BB_1\otimes\BB_2$ of cells we get $\Theta_\BB=\Theta_{\BB_1}\otimes\Theta_{\BB_2}$. We will drop the indices on $\Theta$ in the sequel. Then it is clear from the definition (Lemma~\ref{lem:spp}) that the support algebra construction behaves naturally under global transposition, i.e., we have $\Spp(\Theta\AA,\Theta\BB_1)=\Theta\Spp(\AA,\BB_1)$. Moreover, when $\BB_1$ is a tensor product of cells we get $\Theta\BB_1=\BB_1$. Hence in (\ref{RR2x}) we find for the automorphism $\widetilde\alpha=\Theta\alpha\Theta$:
\begin{eqnarray}
\widetilde\RR_{2x}
   &=& \Spp\Bigl(\Theta\alpha\Theta\bigl(\AA_{2x}\otimes\AA_{2x+1}\bigr),\
                        \bigl(\AA_{2x-1}\otimes\AA_{2x}\bigr)\Bigr)   \nonumber\\
   &=& \Spp\Bigl(\Theta\alpha\bigl(\AA_{2x}\otimes\AA_{2x+1}\bigr),\
                        \Theta\bigl(\AA_{2x-1}\otimes\AA_{2x}\bigr)\Bigr)   \nonumber\\
   &=& \Theta\Spp\Bigl(\alpha\bigl(\AA_{2x}\otimes\AA_{2x+1}\bigr),\
                        \bigl(\AA_{2x-1}\otimes\AA_{2x}\bigr)\Bigr)=\Theta\RR_{2x}.
\nonumber
\end{eqnarray}
Since this has the same dimension as $\RR_{2x}$, we find from (\ref{dimtransfer}) that $\ind(\Theta\alpha\Theta)=\ind\alpha$.
\qed

\proof[of Prop.~\ref{classicalfactor}]
Due to the multiplication formula, we only need to consider the case that $\alpha$ is equal to the identity ($=\beta$) on $\DD$. Consider any operator $A_x\in\AA_x$ in a single cell, and let $D_y\in\DD_y$ be a diagonal element in another cell. Then
$$\alpha(A_x)D_y=\alpha(A_x)\alpha(D_y)=\alpha(A_xD_y)=\alpha(D_yA_xD_y)=D_y\alpha(A_x).$$
Hence the finitely localized element $\alpha(A_x)$ commutes with all diagonal operators on the neighboring sites, and $\alpha(A_x)\in\AA_x\otimes\bigotimes_{y\neq x}\DD_y$. This algebra is best seen as a direct sum of copies of $\AA_x$, labeled by configurations $c=\{c(y)\}$ with $c(y)\in\Alf_y$ of all cells $y\neq x$ in the localization region of $\alpha(\AA_x)$. A homomorphism of $\AA_x$ into this algebra splits into one homomorphism into each summand, which in turn is given by a unitary $U_x(c)$.  Hence we can summarize the action of $\alpha$ on $\AA_x$ as
\begin{equation}
    \alpha(A_x)=\sum_cU_x(c)^*A_xU_x(c)\otimes P(c),
\end{equation}
where $P_c$ is the  minimal projection of the diagonal algebra corresponding to $c$. We also know that diagonal elements of $\AA_x$ are fixed, so $U_x(c)$ is itself diagonal, say $U_x(c)\ket a=u(a,c)\ket a$. This leads to
\begin{equation}\label{localphase}
    \alpha\bigl((\ketbra ab)_x\bigr)=\sum_c\frac{u(b,c)}{u(a,c)}\ (\ketbra ab)_x\otimes P(c).
\end{equation}
The commutation of $\alpha(\AA_x)$ and $\alpha(\AA_y)$ for $x\neq y$ introduces further conditions on the phase functions $u$. But rather than analyzing these in detail, we use the Lemma to conclude directly from Eq.~(\ref{localphase}) that $\ind\alpha=1$. To this end, note that by applying the homomorphism $\alpha$ to (\ref{localphase}) and using $\alpha(P(c))=P(c)$, we get a corresponding formula for the iterate of $\alpha$:
\begin{equation}\label{localphasen}
    \alpha^n\bigl((\ketbra ab)_x\bigr)
       =\sum_c\left(\frac{u(b,c)}{u(a,c)}\right)^n\ (\ketbra ab)_x\otimes P(c),
\end{equation}
for any $n\in\Ir$.
Clearly, the localization region of this operator does not increase with $n$, so by the first part of the Lemma we get $\ind\alpha=1$. Alternatively, we can apply $\Theta$ to the equation, using $\Theta(P(c)=P(c)$. This reverses each of the ketbra operators, so $\Theta\alpha\Theta=\alpha^{-1}$. Hence $\ind\alpha=1$ also follows with the second part of the Lemma.
\qed

An index for classical reversible cellular automata has been defined albeit only in the translationally invariant case. According to G. A. Hedlund (\cite{Hedlund69}, section~14), the definition  is due to L.~R.~Welch, so we will call it the Welch index $i_W$ here. For the definition itself we will follow Kari \cite{Kari_index}, where it is introduced in section~3. We will show that this coincides with the quantum index. Hence the quantum index is a possible extension to non-translationally invariant systems. It is very likely that the theory in \cite{Kari_index} can also be extended directly, but we have not gone to the trouble of checking all the details.

In the translation invariant case, all cell alphabets $\Alf_x\equiv\Alf$ are the same, and a cellular automaton is a map $\phi: \Alf^\Ir\to\Alf^\Ir$. Let $r$ be a ``large enough'' integer, and $R_\phi^r$ the set of $4r$-tuples of the form

$$\Bigl(c(0),\cdots,c(2r-1),(\phi c)(-r),\cdots,(\phi c)(r-1)\Bigr),$$

 where $c$ runs over all infinite configurations. Then the Welch index of $\phi$ is defined as
\begin{equation}\label{iWelsh}
    i_W(\phi)=\frac{|R_\phi^r|}{{|\Alf|}^{3r}}\ .
\end{equation}
Clearly, for the identity only $|\Alf|^3$ different letters occur here, so $i_W(\id)=1$. Similarly, for a shift we get $i_W(S)=|\Alf|$, and parallel application of $\phi$ and $\phi'$ to parallel chains yields $i_W(\phi\times\phi')=i_W(\phi)i_W(\phi')$.

The non-trivial results about the index, and the structure theory of reversible, translationally invariant classical cellular automata is developed in \cite{Kari_index}, with key results analogous to our paper: the expression (\ref{iWelsh}) does not depend on $r$ (provided it is large enough). The product formula holds, and an automaton $\phi$ is locally implementable iff $i_W(\phi)=1$. Moreover, every such automaton can be decomposed into shifts and locally implementable ones. Since a classical ``local implementation'' implies a partitioned representation of the quantum automaton, we can put these facts together to conclude that
\begin{equation}\label{iW=ind}
    i_W(\phi)=\ind\alpha,
\end{equation}
for any quantum cellular automaton which restricts on the diagonal subalgebra to a classical CA given by $\phi$. In this sense our theory is a direct generalization of Kari's work, extended by the aspects of deformation classes (which make no sense in the classical discrete setting) and local computability (which makes no sense in the translation invariant setting).

\subsection{Interlude: More Analogies between Walks and Cellular Automata}
The two definitions of the index, (\ref{defIndW}) for walks and
(\ref{defIndA}) for cellular automata are not directly analogous. Here we would like to point out the differences, and discuss how to make the analogy between these two cases even tighter by supplying the missing analogous definitions.

The definition (\ref{defIndA}) considers a part of the system split into three parts L-M-R. Based on suitable inclusions, it gives a formula for the index, which immediately makes obvious that it is always a positive rational. In contrast, the walk expression Eq.~(\ref{defIndW}) is a difference of numbers which can take arbitrary real positive values, and only one cut of the system is considered. Moreover, (\ref{defIndW}) made it very easy to prove the continuity of the index under deformations, whereas neither (\ref{defIndA}) nor  the abstract considerations of Sects.~\ref{sec:Gimp} and \ref{sec:Glci} clarify continuity for the index of cellular automata. Since continuity is an important feature of our index theory, we will need an appropriate expression also for the automaton case, and the analogies laid out in this subsection are intended to motivate the form of this formula.

\sbsection{Dimension based formula for walks}
Let us first set up an index formula for walks in analogy with (\ref{defIndA}). The analog of the support algebra is the ``support subspace of $\KK_{12}$ in $\KK_2$'',  denoted $\Spp(\KK_{12},\KK_2)$, which is defined for any subspace $\KK_{12}\subset\KK_1\oplus\KK_2$ of the orthogonal direct sum of Hilbert spaces.
Namely, it is the smallest subspace $\LL\subset\KK_2$ such that $\KK_{12}\subset\KK_1\oplus\LL$. Then the analog of Lemma~\ref{sppcomm} holds in the sense that subspaces $\KK_{12}\subset\KK_1\oplus\KK_2$ and
$\KK_{23}\subset\KK_2\oplus\KK_3$ are orthogonal iff $\Spp(\KK_{12},\KK_2)\perp\Spp(\KK_{23},\KK_2)$ are orthogonal. Now consider subspaces $\HH_L\oplus\HH_M\oplus\HH_R\subset\HH$ chosen with the localization constraints as in Sect.~\ref{sec:Defindalpha}. In particular, we require $U\HH_L\perp\HH_R$, and $U\HH_R\perp\HH_L$. Then the direct analog of (\ref{defIndA}) reads
\begin{eqnarray}\label{defIndWW}
    \ind U&=&\dim\Spp(U(\HH_L\oplus\HH_M),\HH_M\oplus\HH_R)-\dim\HH_M
              \nonumber\\
          &=&\rank(P_{MR}UP_{LM})-\dim\HH_M.
\end{eqnarray}
Here the second equality, in which $P_{LM}$ is the projection onto $\HH_L\oplus\HH_M$ etc., follows with
$\Spp(U(\HH_L\oplus\HH_M),\HH_M\oplus\HH_R)=P_{LM}UP_{MR}\HH$.

\proof[of Eq.~(\ref{defIndWW})]
Consider the block matrix for $U$ with respect to the decomposition
$$\HH=\HH_{-\infty}\oplus\HH_L\oplus\HH_M\oplus\HH_R\oplus\HH_{+\infty},$$ where the pieces at the ends contain the appropriate infinite half chains. Using the causality assumptions, we find that
\begin{equation}\label{U-LMR}
    U=\left(\begin{array}{ccc|cc}
       \ast&\ast&0&0&0\\ \ast&\ast&\ast&0&0\\\hline
       0&\fatasterisk&\fatasterisk&\ast&0\\
       0&\hbox{\large\bf0}&\fatasterisk&\ast&\ast\\ 0&0&0&\ast&\ast
       \end{array}\right)\ ,
\end{equation}
where the asterisks stand for any possibly non-zero block. We have highlighted in boldface the block $P_{MR}UP_{LM}$ appearing in (\ref{defIndWW}), and introduced two separating lines, namely
the separation $-\infty,L,M|R,+\infty$ on the domain side (i.e., a vertical line) and the horizontal separation $-\infty,L|M,R,+\infty$ on the range side of $U$. These separators do {\em not} cross on the diagonal, which is why we cannot simply compute the index from the highlighted block via Eq.~(\ref{defIndW}). However, this is easily amended by multiplying with a suitable shift: we can introduce a basis in each $\HH_x$, and hence in $\HH$, effectively making all underlying cell dimensions one-dimensional. In this representation we can introduce a shift operation $S$, and clearly $U'=S^{|M|}U$ will be a unitary with the same matrix elements as $U$ shifted vertically by $|M|=\dim\HH_M$. Obviously, $\ind U'=\ind U+|M|$, which explains the second term in (\ref{defIndWW}), and leaves us with proving that $\ind U'=\rank(P_{MR}UP_{LM})$.
Clearly, for this task the further subdivision of the blocks is irrelevant, and we can consider a general block decomposed unitary operator
$$ U=\left(\begin{array}{c|c}U_{11}&0\\\hline U_{21}&U_{22}\end{array}\right)$$
with a finite rank upper right corner. We have to show that $\rank U_{21}=\tr(U_{21}^*U_{21})$. But from the unitarity equation it follows that $U_{21}$ is a partial isometry, so $U_{21}^*U_{21}$ is its domain projection, whose dimension is indeed the rank of $U_{21}$.
\qed

\sbsection{Half neighborhoods and one-cut dimension formulas}
The proof of formula (\ref{defIndWW}) was essentially by reduction to the case of ``half neighborhoods'', i.e., the case that $[x_-,x_+]=[x,x+1]$, in which no influence ever spreads to the left. Then one of the off-diagonal blocks of the unitary entering (\ref{defIndW}) vanishes, and we saw directly that the other block gives an integer contribution, which can be interpreted as a dimension.

Similarly, for a half-neighborhood cellular automaton we demand $\alpha(\AA_x)\subset\AA_{x-1}\otimes\AA_{x}$, where once again we have chosen the convention to match information traveling to the right, observing the Heisenberg picture. In particular, this condition is satisfied by the shift. For such automata with we can simplify the index formula in a way quite analogous to the case of half-neighborhood walks.
Indeed, setting
\begin{equation}\label{halfstepAlgs}
    \TT_x=\Spp(\alpha(\AA_x),\AA_x) \quad\mbox{and }
    \NN_x=\Spp(\alpha(\AA_x),\AA_{x-1}),
\end{equation}
we can employ the same arguments as in Sect.~\ref{sec:Iauto} to conclude that these commute, and must be full matrix algebras $\TT_x\cong\MM_{t(x)}$ and $\NN_x\cong\MM_{n(x)}$. Then, further following the previous reasoning, $\alpha(\AA_x)=\NN_{x}\otimes\TT_x$ and $\AA_x=\TT_x\NN_{x+1}$. This yields the dimension equation
\begin{equation}\label{halfstepDims}
    d(x)=n(x)t(x)=t(x)n(x-1).
\end{equation}
Hence the integers $n(x)$ do not depend on $x$. In fact, $\ind\alpha=n(x)$. This follows readily from the observation that
$$ \RR_{2x+1}= \Spp\Bigl(\alpha\bigl(\AA_{2x}\otimes\AA_{2x+1}\bigr),\
                        \bigl(\AA_{2x+1}\otimes\AA_{2x+2}\bigr)\Bigr)
           =\TT_{2x+1}\otimes\idty_{2x+2}.
$$
Therefore, from Eq.~(\ref{dimtransfer}) we get
$$\ind\alpha=d(2x+1)/r(2x+1)=d(2x+1)/t(2x+1)=n(2x+1).$$

In fact, for any cell structure on which a shift is available, we could have used this slightly simplified setup to define the index of any $\alpha$ by first shifting and regrouping to a half-neighborhood automaton, and correcting by a factor depending on the size of the necessary shift. However, since such a shift is not available in general, there was no gain in this approach.

For nearest neighbor automata the structure of support algebras using just a single cut, i.e., algebras of the form $\Spp(\alpha(\AA_L),\AA_R)$ are not sufficient to define the index. As a counterexample, consider
a unitary evolution $A\mapsto U^*AU$ for $U\in\AA_L\otimes\AA_R$ with $\AA_L=\AA_R=\MM_d$. Clearly, as a locally implementable operation, this has always trivial index. But the support algebra written above can be just about anything. For example, taking a ``controlled unitary'' $U=\sum_i \kettbra i\otimes U_i$, we have
$\Spp(\alpha(\AA_R),\AA_L)$ as the algebra of diagonal matrices, and $\Spp(\alpha(\AA_L),\AA_R)$ generated by the $U_i^*U_j$, which for $\AA_R\cong \MM_r$ can easily be  $\AA_R$. On the other hand, one can easily construct an automorphism with the same support algebras, but index $d$.

The algebraic structure of support algebras across a single cut is also insufficient for us in another way: it is a discrete structure, hence does not go to any trivial value as we deform an $\alpha$ to the identity. Therefore, if only to get the continuity of the index we need to look at some continuously varying quantities. We found a formula with just these properties by looking at cellular automaton analogs of the simple walk formula (\ref{defIndW}). The following section is devoted to the proof of this formula.

\subsection{One-cut quotient formula}
\sbsection{Notation for normalized traces}
A continuously varying quantity measuring the difference between subalgebras and going to a trivial value as they come to coincide may be some kind of angle, or overlap, between the linear subspaces. It is natural to measure such angles with respect to the only scalar product between algebra elements, which is canonically defined in our context. Indeed, let $\tau$ denote the {\em normalized trace} on the algebra $\AA$ of the entire chain. That is, on any matrix subalgebra $\BB\subset\AA$, $\BB\cong\MM_d$, we define $\tau(A)=\frac1d\tr(A)$, where $\tr$ is the usual matrix trace on $\MM_d$, which is $1$ on minimal projections. The reason for this normalization is that in contrast to the matrix trace, the value of $\tau$ does not change if we consider $A$ to be embedded as $A\otimes\idty$ in some larger subalgebra $\BB\otimes\BB_1$, and is hence a well defined state on chain algebra $\AA$. In fact, it is the unique  state (normalized positive functional) on $\AA$ with the property that $\tau(AB)=\tau(BA)$.  This characterization is purely algebraic, which implies that, for every automorphism $\alpha$ of $\AA$ and any $A\in\AA$, we have $\tau(\alpha(A))=\tau(A)$.

We now define the scalar product between algebra elements by
\begin{equation}\label{HSproduct}
    \braket xy=\tau(x^*y).
\end{equation}
The completion of the algebra $\AA$ as a Hilbert space with this scalar product is called the GNS-Hilbert space $\HH_\tau$ of the tracial state $\tau$. We write $\ket x\in\HH_\tau$ for the vector obtained by embedding $x\in\AA$ in the completion. The trace of operators on $\HH_\tau$ will be denoted by $\ttr$, to avoid confusion with the matrix trace $\tr$ of some elements of $\AA$, which is also used later.
Since $\tau$ is preserved by any automorphism $\alpha$, we can define a unitary operator $V_\alpha$ on $\HH_\tau$ with
\begin{equation}\label{Valpha}
    V_\alpha\ket x=\ket{\alpha(x)}.
\end{equation}

Consider now a finite or infinite dimensional subalgebra $\BB\subset\AA$. By $P$ we denote the orthogonal projection onto the closure of $\BB$ in $\HH_\tau$. For finite matrix algebras $\BB\cong\MM_d$ the matrix units $e_{ij}\cong\ketbra ij$ clearly form a basis $\BB$, and one readily verifies that
$\{\sqrt d\,\ket{e_{ij}}\}_{i,j=1}^d$ is an orthonormal basis of this $d^2$-dimensional subspace. It is sometimes also helpful to represent the projection $P$ as an integral over unitaries. That is, for a finite dimensional subalgebra $\BB\subset\AA$ we have
\begin{equation}\label{twirl}
    P=d^2 \int\!\!dU\ \kettbra U,
\end{equation}
where the integral is over the unitary group of $\BB$, and $dU$ denotes the normalized Haar measure.
For infinite dimensional subalgebras such formulas are not available. We will only need infinite dimensional projections of this type for half chain algebras, which are the closure of an increasing net of finite dimensional subalgebras $\BB_n$. In this case the family of projections $P_n$ associated with the approximating algebras is also increasing, and converges strongly to $P$.

When $\BB_1\subset\AA$ and $\BB_2\subset\AA$ are commuting matrix subalgebras, the corresponding matrix units, say $e_{ij}^{(1)}$ and $e_{ab}^{(2)}$, can be taken together as a set of matrix units for $\BB_1\otimes\BB_2$, and we get
$$ \braket{e_{ij}^{(1)}}{e_{ab}^{(2)}}=\tau({e_{ji}^{(1)}}{e_{ab}^{(2)}})
   =\frac1{d_1d_2}\delta_{ij}\delta_{ab}.$$
Therefore we get
\begin{equation}\label{overlap1}
  P_1P_2=\kettbra\idty.
\end{equation}
This equation also holds for infinite pieces of the chain, e.g., a right and a left half chain localized
on disjoint subsets of $\Ir$. This readily seen by approximating each half chain by finite matrix algebras and using the strong convergence of projections. Note that (\ref{overlap1}) also implies that $P_1$ and $P_2$ commute, and the projections $P_i-\kettbra\idty$ with the intersection removed are orthogonal.

\sbsection{Overlap of algebras}
Of course, if two algebras do not commute, which requires that the localization regions of $\BB_1$ and $\BB_2$ overlap, the geometric position of the subspaces $P_i\HH_\tau$ is not so simple. Even if the algebras have trivial intersection $\BB_1\cap\BB_2=\Cx\idty$, so that $(P_1\HH_\tau)\cap(P_2\HH_\tau)=\Cx\ket\idty$ the vectors in the remainder can now have angles different from $\pi/2$, and may even approximate each other. This leads to the following definition of a quantitative notion of the overlap of two algebras, which will be the basis of the index formula we develop in this section. We state it together with a few of its basic properties. A matrix algebra here always means a subalgebra, which is isomorphic to $\MM_d$ for some $d<\infty$ and contains the identity of $\AA$.

\begin{lem}\label{overlemma}
For any to subalgebras $\BB_1,\BB_2\subset\AA$, with corresponding orthogonal projections $P_1,P_2$ on $\HH_\tau$,  we define their {\bf overlap} as
\begin{equation}\label{overlap}
    \ovlp(\BB_1,\BB_2)=\sqrt{\ttr(P_1P_2)}\ \in[0,\infty].
\end{equation}
Then \begin{enumerate}
\item $\ovlp(\BB_1,\BB_2)=1$, for commuting matrix algebras
\item $\ovlp(\BB_1,\BB_2)\geq1$, for any two subalgebras (with unit).
\item Let $\BB_1,\BB_2\subset\BB_{12}$ and $\BB_3,\BB_4\subset\BB_{34}$ are matrix subalgebras such that $\BB_{12}$ and $\BB_{34}$ commute. Then
\begin{equation}\label{overlaptensor}
    \ovlp(\BB_1\BB_3,\BB_2\BB_4)=\ovlp(\BB_1,\BB_2)\ovlp(\BB_3,\BB_4).
\end{equation}
\item Let $\BB_1,\BB_2,\BB_3,\BB_4$ be a chain of matrix algebras such that $[\BB_i,\BB_j]=\{0\}$, except for $i,j=2,3$. Then
    $\ovlp({\BB_1\BB_2,\BB_3\BB_4})=\ovlp({\BB_2,\BB_3})$.
\item Let $\BB_1,\BB_2,\BB_3$ be commuting matrix algebras with $\BB_2\cong\MM_d$. Then
    $\ovlp(\BB_1\BB_2,\BB_2\BB_3)=d$.
\item Let $\BB_1\cong\MM_d$ and $\BB_2$ be finite dimensional matrix algebras, and $\alpha$ an automorphism of the ambient algebra such that $\norm{\alpha(x)-x}\leq\varepsilon\norm x$ for $x\in\BB_1$.
    Then $$ \abs{\ovlp(\alpha(\BB_1),\BB_2)-\ovlp(\BB_1,\BB_2)}\leq \varepsilon d^2.$$
\end{enumerate}
\end{lem}

\proof
Item 1 follows directly from Eq.~(\ref{overlap1}). For item 2, note that $P_i\geq\kettbra\idty$. Since we are only considering subalgebras containing the identity of the ambient algebra, the parenthetical remark is only added for emphasis.  Item 3 follows by observing that the
normalized trace on $\BB_{12}\BB_{34}\cong\BB_{12}\otimes\BB_{34}$ is fixed to be the product of normalized traces on the subalgebras. In this tensor product representation one readily verifies that the projection belonging to $\BB_1\BB_3$ is the tensor product $P_1\otimes P_3$, and similarly for $P_2,P_4$. Then the formula follows because the trace $\ttr$ also factorizes. \\
Finally item 4 follows by taking $\BB_1\BB_4$ and $\BB_2\BB_3$ as the pairing of item 2, and using item 1 to conclude that $\ovlp(\BB_1,\BB_3)=1$.

To prove the continuity estimate in item 6, note that overlaps are $\geq1$, so that by a gradient estimate on the square root function
$$ \abs{\ovlp(\alpha(\BB_1),\BB_2)-\ovlp(\BB_1,\BB_2)}
    \leq\frac12\left|\ttr((\widetilde P_1-P_1)P_2)\right|
    \leq\frac12\norm{\widetilde P_1-P_1}_1,$$
where $\widetilde P_1,P_1,P_2$ are the projections belonging to $\alpha(\BB_1),\BB_1,\BB_2$, and
$\norm X_1$ denotes the trace norm of $X$. We use the representation of $P_1$ in the form (\ref{twirl}), so that
$$ \widetilde P_1-P_1=d^2\int\!\!dU\ \Bigl(\kettbra{\alpha(U)}-\kettbra U\Bigr).$$
We will estimate the trace norm of this expression by estimating the integrand, and using that, for any unit vectors $\phi,\psi$ in a Hilbert space, $\norm{\,\kettbra\phi-\kettbra\psi\,}_1\leq2\norm{\phi-\psi}$. Indeed, for unitaries like $U$ and $\alpha(U)$ the vectors $\ket U$ and $\ket{\alpha(U)}$ have norm 1 in $\HH_\tau$. Moreover, $\norm{\ket{\alpha(U)-U}}^2=\tau\Bigl((\alpha(U)-U)^*(\alpha(U)-U)\Bigr)\leq \varepsilon^2\norm U\tau(\idty)=\varepsilon^2$. Hence
$$\frac12\norm{\widetilde P_1-P_1}_1\leq \frac{2d^2}2\varepsilon\int\!\!dU=d^2\varepsilon.$$
\qed

\sbsection{Index formula}
With these preparations we can state the main result of this section:

\begin{prop}\label{davidsformula}
Let $\AA_<=\AA_{(-\infty,0]}$ and $\AA_>=\AA_{[1,\infty)}$, and $\AA_L,\AA_R$ the algebras of any two neighboring cell algebras for a nearest neighbor grouping of the chain.   Then
\begin{equation}\label{indDavid}
    \ind(\alpha)=\frac{\ovlp\Bigl({\alpha(\AA_<),\AA_>}\Bigr)}{\ovlp\Bigl({\alpha(\AA_>),\AA_<}\Bigr)}
               =\frac{\ovlp\Bigl({\alpha(\AA_L),\AA_R}\Bigr)}{\ovlp\Bigl({\alpha(\AA_R),\AA_L}\Bigr)}.
\end{equation}
Moreover, if $t\mapsto\alpha_t$ is a strongly continuous family of cellular automata with the same cell structure and neighborhood scheme, then $\ind(\alpha_t)$ is constant.
\end{prop}

\proof The square of the numerator of the second index expression is $\ttr V_\alpha P_LV_\alpha^*P_R$, and we first verify that this expression is unchanged if we choose a larger $L$ and $R$, say $L'=L_1\cup L$ and $R'=R\cup R_1$. Indeed, the algebras $\alpha(\AA_{L_1}), \alpha(\AA_L),\AA_R,\AA_{R_1}$ satisfy the conditions of Lemma~\ref{overlemma}, item 4. Moreover, arguing as for formula (\ref{overlap1}) we see that not only the trace, but even the operator is independent of an enlargement. Hence taking a strong limit we obtain the corresponding expression for the infinite half-chains.

From Lemma~\ref{overlemma}, item 3, it is clear that the expression thus defined satisfies the tensor product property in fact, numerator and denominator do so independently. Moreover, the formula is valid for shift automorphisms by virtue of item 5 of the Lemma.

Now let $\sigma$ be a tensor product of shifts with $\ind\sigma=\ind\alpha$, and let $\ind'\alpha$ be the value the formula in the proposition gives for any automorphism $\alpha$. Then we have $\ind'(\alpha\otimes\sigma^{-1})=\ind'(\alpha)(\ind\sigma)^{-1}=\ind'\alpha/(\ind\alpha)$. So it remains to prove that $\ind'\alpha=1$ for every $\alpha$ with $\ind\alpha=1$, which by Thm.~\ref{thmIndA} means that $\alpha$ is implemented by two layers of block unitaries. Moreover, we can forget all unitaries acting only on one side of the separation, since they do not change the algebras. Only one unitary $U_{LR}\in\AA_L\otimes\AA_R$ connecting the cells immediately adjacent to the cut remains. Hence it only remains to prove the Lemma below.

For the continuity statement observe that we can make a nearest neighbor grouping jointly for all $\alpha_t$, so we can apply Lemma~\ref{overlemma}, item 6, to see that both denominator and numerator of the second fraction in (\ref{indDavid}) are continuous. Moreover, the denominator is $\geq1$, so $\ind(\alpha_t)$ is a continuous function of $t$. On the other hand, numerator and denominator of $\ind\alpha$ have to divide every cell dimension, so there is only a finite choice of possible values, given the neighborhood structure. Hence  $\ind(\alpha_t)$ is constant.
\qed

In order to state the remaining Lemma, let us consider any automorphism $\alpha$ of a tensor product $\AA_L\otimes\AA_R$ of two finite dimensional matrix algebras. We can take $\AA_x = \BB(\HH_x)$, for some
$d_x$-dimensional Hilbert spaces $\HH_x$). In that case, the
automorphism $\alpha$ is implemented by conjugation with a unitary:
$\alpha(x) = U x U^*$, for some $U\in U(\HH_L\otimes \HH_R)$. We will
express the fraction appearing in (\ref{indDavid}) in terms of $U$.

In the following calculation, it turns out to be convenient to
introduce orthonormal bases in the Hilbert spaces involved. Denote
these bases by $\{\ket i\}\subset \HH_L$ and $\{\ket a\}\subset \HH_R$
respectively. These bases allow us to define the notion of a
\emph{partial transpose} $U^\Gamma$ of $U$. We set
$$
	\bra{ia}(U^\Gamma)\ket{jb} = \bra{ja}U\ket{ib}.  $$
While the partial transpose depends on the basis used to define it,
one may easily convince oneself that the expression appearing in next lemma is independent of that choice.

\begin{lem}\label{lem:contraction}
	When $\alpha(x)=UxU^*$ is an automorphism of the tensor product $\AA_L\otimes\AA_R$ of finite
    dimensional  matrix algebras, we have
	\begin{equation}\label{eqn:partialTranspose}
	 \ovlp(\alpha(\AA_L),\AA_R)^2=
		\frac1{d_Ad_B}\tr\left(\bigl(U^\Gamma U^{\Gamma*}\bigr)^2\right)
	\end{equation}
	Moreover, this expression is invariant under the substitution
	$\AA_L \leftrightarrow \AA_R$.
\end{lem}

\proof
We introduce matrix units $e_{ij}=\ketbra ij$ as above and use that
the $\sqrt d\,{e_{ij}}$ form an orthonormal basis of $\AA_L$. Thus the projection $P_L$ on $\HH_\tau$
corresponding to this algebra is $P_L=d_L\sum_{ij}\kettbra{e_{ij}}$. Of course, the transformed projection $\widetilde P_L$ for $\alpha(\AA_L)$ is obtained by substituting $\ket{\alpha(e_{ij})}$ for $\ket{e_{ij}}$. Similarly, we set $f_{ab}=\ketbra ab$ and, accordingly, $P_R=d_R\sum_{ab}\kettbra{f_{ab}}$.
Then the left hand side of (\ref{eqn:partialTranspose}) becomes
$$
\ovlp(\alpha(\AA_L),\AA_R)^2=\ttr(\widetilde P_LP_R)
   = d_Ld_R\sum_{ijab}\left|\braket{f_{ab}}{\alpha(e_{ij})}\right|^2,
$$
with the scalar products
\begin{eqnarray}
\braket{f_{ab}}{\alpha(e_{ij})}
  &=& \sum_{kc}\tau(e_{kk}f_{ba} Ue_{ij}f_{cc}U^*)   \nonumber\\
  &=& \frac1{d_Ld_R}\sum_{kc}\bra{ka}U\ketbra{ic}{jc}U^*\ket{kb} \nonumber\\
  &=& \frac1{d_Ld_R}\sum_{kc}\bra{ia}U^\Gamma\ketbra{kc}{kc}U^{\Gamma*}\ket{jb} \nonumber\\
  &=& \frac1{d_Ld_R}\bra{ia} U^\Gamma U^{\Gamma*}\ket{jb}\nonumber\\
\end{eqnarray}
Altogether we get Eq.~(\ref{eqn:partialTranspose}).

The second claim becomes a simple corollary: we must verify that the right hand side is unchanged by the substitution $U\mapsto U^*$. Since the adjoint operation commutes with the partial transpose, this amounts to a cyclic rearrangement under the trace.
\qed

\section{Outlook}
\label{sec:out}
Two directions of generalization of the theory presented here are especially suggestive and are, in fact, the subject of ongoing work in our group. We briefly comment on the prospects.

\subsection{Approximate causality}
In many situations in physics causality is only approximately satisfied, e.g., as a bound $\norm{U_{xy}}\leq f(|x-y|)$ for some function $f$ going to zero at infinity. For example, the unitary groups generated by finite range Hamiltonians would satisfy this, but are never strictly causal in the sense required here. For the key Lemma.~\ref{finrank} powerful generalizations exist \cite{Seiler}. From these it is clear that the index of approximately causal unitaries is definable, integer valued, satisfies a product formula, and is zero for the unitaries arising from Hamiltonian subgroups. The part of the theory which is less clear is related to the converses, namely the construction of strictly causal unitary walks with the same index, approximating a given approximately causal unitary.

In the case of cellular automata the right notions of approximate causality are not clear. Ideally, one would only demand that $\alpha$ be an automorphism of the quasi-local algebra \cite{BraRo}. By definition, this means that the image of any localized element can be approximated in norm by localized ones. The idea of support algebras (which largely relies in its present form on picking a finite basis) is certainly too simplistic, and in any case it is unlikely that such algebras will always turn out to be finite dimensional matrix algebras. One may speculate, whether the index should take on also irrational values, but this seems unlikely, because of its dependence on the cell structure: for chains of homogeneous cell dimension $2$, the index is always a power of $2$, and not a dense set of rationals.

\subsection{Higher lattice dimension}
As we have shown, in lattice dimension 1 three possible classifications of walks and automata coincide:
(1) the classification by locally computable invariants, (2) the classification modulo locally implementable,  unitaries, and (3) the classification up to homotopy. It is very unlikely that these three coincide in higher dimension as well. In the one-dimensional theory we could allow the local systems to grow, but also the localization regions. In the higher dimensional case we will use a translation invariant metric to bound the neighborhood sizes of a ``local'' system. Generalizations can be built on coarse geometry \cite{coarse}.

Locally computable invariants are probably trivial. For example, an arbitrarily large patch of the shift automorphism can be connected to the identity outside a finite enlargement of the patch. In this sense the shifts have the same invariants as the identity. To get a more interesting theory, one should take other regions for the definition of ``locally'' computable, e.g., computability on cones \cite{Buchholz}, or computability {\em outside} of any arbitrarily large region.

The classification modulo local implementability is especially interesting from the physical point of view, but it might turn out to be rather wild. For example, consider some self-intersection free path in the lattice, which comes from infinity and goes to infinity. Fix walks/cellular automata, which are equal to the identity off this path and allow any of the one-dimensional systems along this path. As long as we fix the path, we can apply the one dimensional theory. For two paths, which keep a finite distance from each other, it is easy to envisage local swap-type unitaries taking one path to the other, which would bring the corresponding path-related indices under the same roof. However, if the paths move away from each other, there will be no such local operation connecting them, so systems with non-trivial indices along these paths fall into different equivalence classes modulo local implementability. However, the equivalence classes of paths modulo ``keeping a finite distance'' are a rather unmanageably large set. A useful classification cannot be expected. Incidentally, the same class of examples shows that the ``invariants computable outside any finite region'' will give a wild set.

For the homotopy classification of walks there is already a theory, based on the K-theory of C*-algebras and its connection with coarse geometry. Indeed, the $K_1$ group of a C*-algebra just classifies the connected components of its automorphism group. This theory will most naturally apply to approximately causal walks, since strict causality cannot even be stated simply in terms of the C*-algebra of the whole system. The connection with coarse geometry is being explored, for example, by Ralf Meyer and his group in G\"ottingen. Surprisingly (to us) it turns out that using Bott periodicity one can see that the $K_1$ group of approximately causal walks alternates between $\Ir$ in odd dimension and $0$ in even dimension. Unfortunately, this theory does not apply readily to the case of cellular automata.

\subsection{Higher dimensional translation invariant systems}
In order to tame the wildness indicated in the previous subsection, one can restrict attention to translationally invariant systems. Immediately, the index of walks gets an obvious definition. The Fourier transform $\widehat U$ of the walk is now a Laurent polynomial in the variables $\exp(ip_k)$, where $(p_1,\ldots,p_s)$ is the momentum vector. Then $\det\widehat U$ is an invertible polynomial and we again conclude $\det\widehat U(p)=\exp(i\sum_kn_kp_k)$ for some integers $n_k$. The lattice vector $(n_1,\ldots,n_s)$ can be called the index, in direct generalization of the one-dimensional case.

For cellular automata, which are products of partial shifts and local block unitaries, we can just define the index via the shift content contained in such a representation, obtaining some vector with rational components. Since it is not known, whether any QCA is of this form, this is very unsatisfactory. As a step in the right direction, one can define an index by reduction to the 1D case, without using any special decomposition:
Suppose we choose vectors $a_1,\ldots,a_{n-1}$ in an $n$-dimensional lattice, and we identify sites differing by integer multiples of these vectors. Then if the $a_k$ are large enough with respect to the interaction length of the automaton, this gives a well-defined cellular automaton evolution $\alpha[a_1,\ldots,a_{n-1}]$ on the quotient lattice, which now has only one unbounded direction, so we can assign an index to it. Then we call a vector $q\in\Rl^n$ the index of $\alpha$, if
\begin{equation}\label{indQuotienTI}
    \ind \alpha[a_1,\ldots,a_{n-1}]=\det[q,a_1,\ldots,a_{n-1}],
\end{equation}
where on the right hand side the square bracket denotes the matrix with the specified column vectors. It can be verified easily that this gives the result indicated before, for any cellular automaton, which has a decomposition into partial shifts and partitioned unitaries. But in general it is not even clear that the left hand side must depend linearly on each $a_k$.

\section*{Acknowledgments}
We gratefully acknowledge the support of the DFG (Forschergruppe 635) and the EU (projects CORNER and QUICS), as well as the Erwin Schr\"odinger Institute.

\bibliography{qciLit} 
\bibliographystyle{abbrv}

\end{document}